\documentclass[%
aip,
amsmath,amssymb,
reprint,%
author-year,%
nofootinbib
]{revtex4-2}

\usepackage{graphicx,array,booktabs,rotating}
\usepackage{dcolumn}
\usepackage{bm}
\usepackage{mathptmx}
\usepackage[english]{babel}
\usepackage{float,color,xcolor}
\usepackage{hyperref}
\usepackage{etoolbox} 
\usepackage{tabularx,tabulary}
\usepackage[capitalize]{cleveref}
\usepackage{times}
\usepackage{multirow}
\usepackage{textcomp,fontenc}
\hypersetup{
	colorlinks,
	linkcolor={blue!100!black},
	citecolor={blue!100!black},
	urlcolor={blue!100!black}
}
\newcommand{\PRLsep}{\noindent\makebox[\linewidth]{\resizebox{0.3333\linewidth}{1pt}{$\bullet$}}\bigskip}
\makeatletter
\renewcommand*{\fnum@figure}{{\normalfont\bfseries \figurename~\thefigure}}
\renewcommand*{\@caption@fignum@sep}{\textbf{. }}
\renewcommand*{\fnum@table}{{\normalfont\bfseries \tablename~\thetable}}
\renewcommand*{\@caption@fignum@sep}{\textbf{. }}
\makeatother

\usepackage{chngcntr,xr,titlesec}
\titleformat{\subsubsection}[runin]
{\sffamily\small\bfseries}{\thesubsubsection.}{0.25em}{}[:]
\titlespacing{\subsubsection}{0pc}{1.5ex plus .1ex minus .2ex}{0.25em}
\usepackage[page]{totalcount}
\usepackage{etoolbox,fancyhdr,xcolor}%
\renewcommand{\headrule}{\vbox to 5pt{\hbox to\headwidth{\hrulefill}\vss}}
\makeatletter
\patchcmd{\NAT@citex}
{\@citea\NAT@hyper@{%
		\NAT@nmfmt{\NAT@nm}%
		\hyper@natlinkbreak{\NAT@aysep\NAT@spacechar}{\@citeb\@extra@b@citeb}%
		\NAT@date}}
{\@citea\NAT@nmfmt{\NAT@nm}%
	\NAT@aysep\NAT@spacechar\NAT@hyper@{\NAT@date}}{}{}
\patchcmd{\NAT@citex}
{\@citea\NAT@hyper@{%
		\NAT@nmfmt{\NAT@nm}%
		\hyper@natlinkbreak{\NAT@spacechar\NAT@@open\if*#1*\else#1\NAT@spacechar\fi}%
		{\@citeb\@extra@b@citeb}%
		\NAT@date}}
{\@citea\NAT@nmfmt{\NAT@nm}%
	\NAT@spacechar\NAT@@open\if*#1*\else#1\NAT@spacechar\fi\NAT@hyper@{\NAT@date}}
{}{}

\makeatother
\usepackage[pagewise,mathlines]{lineno} 

\begin{document}
	\preprint{AIP/123-QED}	
	\title[\textbf{Journal to be decided} (2022) $|$  Rough draft of the manuscript]{Unsteadiness in hypersonic leading-edge separation}

    \author{S. K. Karthick}
	\email{skkarthick@ymail.com}
	\affiliation{Technion Wind Tunnels Complex, Faculty of Aerospace Engineering, Technion-Israel Institute of Technology, Haifa-3200003, Israel}%

	\author{Soumya R. Nanda}
	\affiliation{Technion Wind Tunnels Complex, Faculty of Aerospace Engineering, Technion-Israel Institute of Technology, Haifa-3200003, Israel}%
	
	\author{J. Cohen}
	\affiliation{Technion Wind Tunnels Complex, Faculty of Aerospace Engineering, Technion-Israel Institute of Technology, Haifa-3200003, Israel}%
	
	\date{\today}
\begin{abstract}
Hypersonic leading-edge separation is studied towards understanding the varying shock-related unsteadiness with freestream Reynolds number ($1.66 \times 10^5 \leq Re_D \leq 5.85 \times 10^5$) in the newly constructed hypersonic Ludwieg tunnel (HLT) at a freestream design Mach number of $M_\infty=6.0$. An axisymmetric flat-face cylinder of base body diameter $D=35$ mm is fitted with protrusions of different fineness ($d/D=0.1,0.2,0.26,0.34$ at $L/D=1.4$) and slenderness ($L/D=0.7,1,1.4,1.9$ at $d/D=0.2$) ratio to induce a wide range of leading-edge separation intensities. Qualitative and quantitative assessments are made using schlieren imaging, planar laser Rayleigh scattering, and unsteady pressure measurements. A well-known to-and-fro shock motion called pulsation and a flapping shock-shear layer oscillation is observed as $Re_D$ changes. A shorter protrusion length ($L/D=0.7$) produces a pressure loading that is four orders higher than the cases with longer protrusion lengths. There exists a critical separation length ($L/D \geq 1.4$) beyond which the separated shear layer trips to turbulence and introduces fluctuations in the recirculation region as $Re_D$ increases. The effect of the separated turbulent shear layer is dampened by an order provided the reattachment angle is shallow by increasing the fineness ratio ($d/D=0.4$). There also exists a critical geometrical parameter ($L/D=1, d/D=0.2$) for which the unsteady modes switch between successive runs based on the upstream fluctuations. From the modal analysis of the Rayleigh scattering images, the first four dominant modes that drive flapping are identified as translatory flapping, sinuous flapping, large and small-scale shedding.
\end{abstract}


\maketitle

\section{Introduction} \label{sec:intro}

\begin{figure*}
  \centerline{\includegraphics[width=0.7\textwidth]{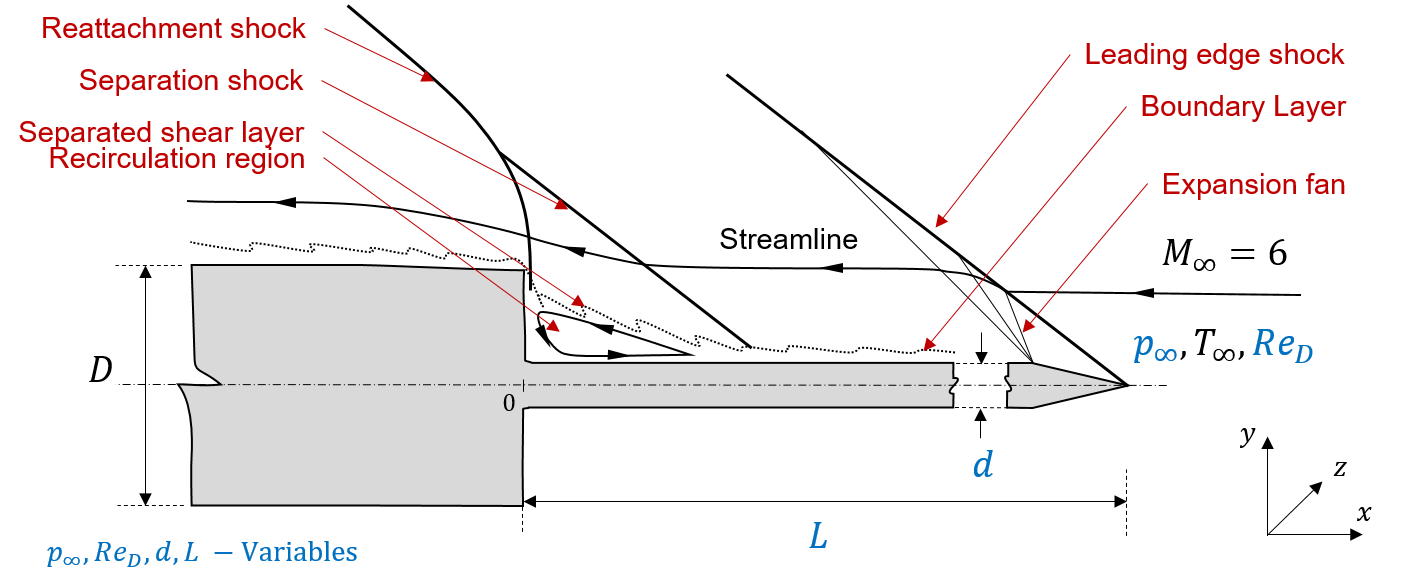}}
  \caption{Key flow features observed in a typical leading-edge separated hypersonic flow over a generic flat-face forebody with an axial protrusion. Prominent flow attributes include leading-edge shock, reattachment shock, separated shear layer, boundary layer, expansion fan, and a streamline passing through the system of shocks and expansion fan. The potential parameters that can affect the occurrence of the aforementioned events include the freestream Reynolds number ($Re_D$) and the geometrical parameters of the protrusion (length-$L$ and diameter-$D$).}
\label{fig:flow_schematic}
\end{figure*}

Hypersonic vehicles being the current state-of-the-art, imposes lot of challenges towards designing the structures which in turn relies on in-depth understandings of the flow evolution and characterization. High-speed vehicles flying at hypersonic speed is often exposed to severe pressure loading and aerodynamic heating due to presence of bow shock ahead of it. Therefore, continuous effort is being made to mitigate the serious threat posed by the immense amount of pressure and heat flux imposition on the vehicles, thereby enhancing the efficacy in accomplishing the objectives. Various ways of active and passive control techniques \citep{Ahmed2011} are being adopted such as energy deposition, opposing jet, cavities, axial protrusions, and combinations of them. Application of axial protrusions in the form of spikes \citep{Maull1960584,Kenworthy1978} is considered as the simplest. The passive control axial protrusion module is used in a variety of missiles and air intakes\citep{Venukumar2006,Kulkarni2008} of high-speed flying vehicles\citep{Sekar2020,Devaraj2020} owing to its low drag benefits. The protrusion tip can take various shapes ranging like sharp-tip, hemisphere, aerodisk, circular, elliptical and so on. Typically, the incorporation of axial protrusion, weakens the strength of the oblique shock by pushing it away from the object through shock-shock and shock-shear layer interaction. Nevertheless, the leading-edge separation on the protrusion leads to severe shock and shear layer interaction and leads to global unsteadiness. Typical flow features seen during the leading-edge separation is shown in Figure \ref{fig:flow_schematic}. Formation of a leading-edge shock is usually evident at the protrusion's tip followed by the formation of a recirculation region due to an adverse pressure gradient. The recirculating zone is accompanied by the shock and separated shear layer formation. Further, a reattachment shock forms near the model shoulder, to turn the flow parallel to the model.  

Depending on the freestream conditions, and the ratio between the protrusion's length and diameter to the base body diameter ($L/D$, $d/D$), different forms of unsteadiness occur. It can be either strong as in the case of pulsation or moderately weak as in oscillation/flapping  \citep{Feszty2004a,Feszty2004b,Sugarno2022}. Typically, in pulsation mode the separated shock from the tip undergoes continuous change in shock shape from conical to hemispherical during its to-and-fro motion along the protrusion's length. Whereas, in flapping, local lateral movement of shock and shear layer is seen. Understanding this unsteadiness is of primary importance in reducing the fluid-structure interactions. Since, the inception of axial protrusions on aerodynamic bodies, experimental and numerical studies were conducted to describe the driving potential behind sustaining the pulsating and flapping mode of unsteadiness. Several theories have also been proposed to assess the flow characteristics pertaining to these conditions. Protrusions of different geometrical features produce different flow physics \citep{Panaras200969}. Recently, \cite{sahoo_2020,sahoo_2021} studied the underlying physical aspects involving shock-related unsteadiness at supersonic speeds. However, information related to hypersonic speeds, particularly for a wide range of Reynolds numbers ($Re_D$), is limited \cite{Ahmed2011}. An additional challenge is designing and constructing facilities suited for studying such flow problems.

\begin{figure*}
  \centerline{\includegraphics[width=0.65\textwidth]{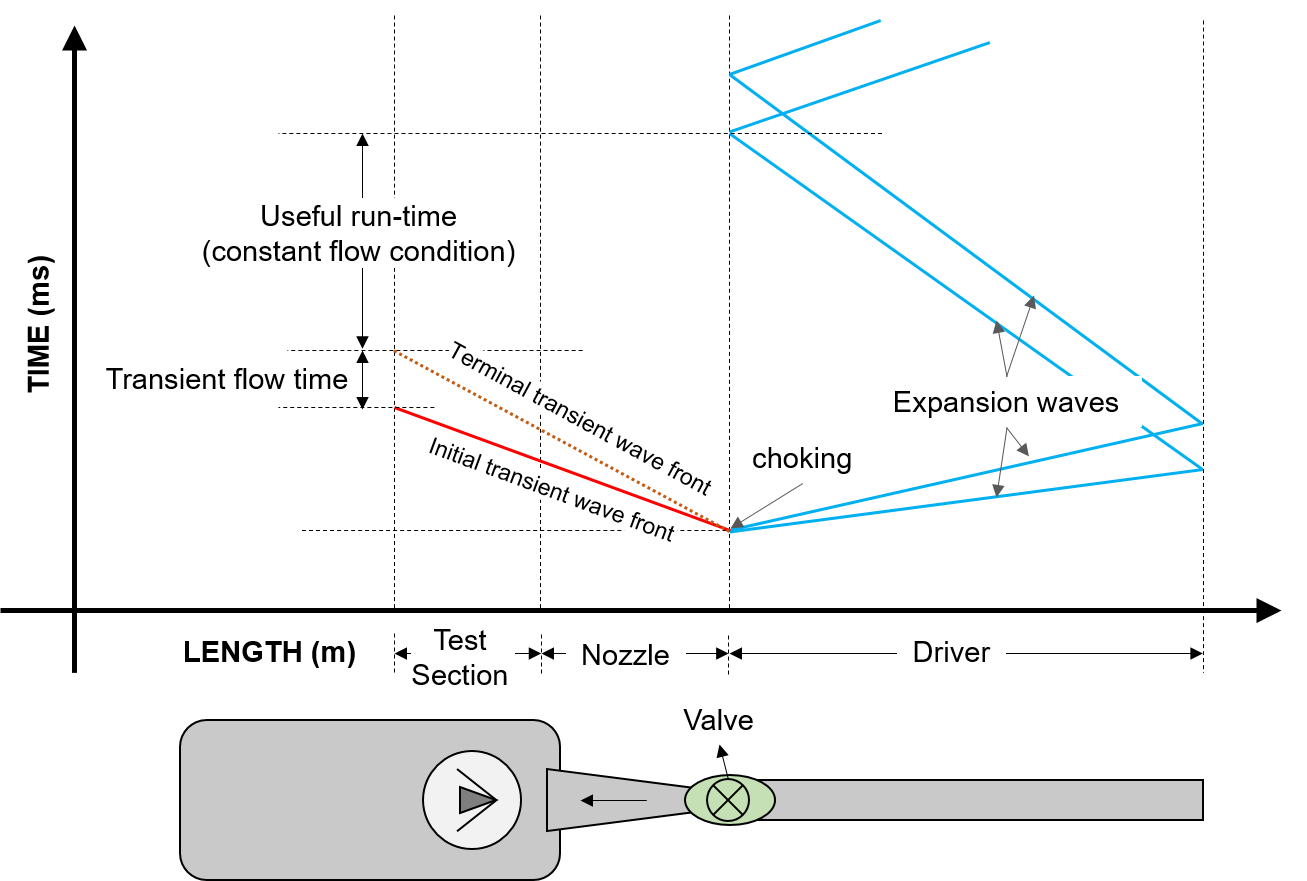}}
  \caption{Typical $x-t$ plot showing the propagation of weak compression waves (right running) and expansion waves (left running) from the point of valve opening ahead of the nozzle and the test section in a hypersonic Ludwieg tunnel mode of operation. The resulted flow go from left to right. The useful run-time is indicated between the time of arrival of terminal transient and reflected expansion waves.}
\label{fig:xt_Ludwieg_tube}
\end{figure*}

In order to generate hypersonic flow with different $Re_D$ at a specified $M_\infty$, hypersonic Ludwieg tunnel can be used, as it is easy to setup, maintain, and control. It comprises of a long storage tube followed by a fast-acting valve, convergent divergent (C-D) nozzle, and dump tank. Typical $x-t$ diagram describes the position of different flow features with respect to time upon valve opening in Figure \ref{fig:xt_Ludwieg_tube}. As the valve opens, expansion waves propagate upstream and simultaneously the test gas gets expanded in the C-D nozzle, resulting in a hypersonic flow depending on the nozzle's area ratio. Until, the expansion fan gets reflected from the end wall and reaches the valve location, useful run time is established in the test section. Ludwieg tunnels can generate cold flows corresponding to a low enthalpy (0.3 to 0.7 MJ/kg) and a relatively low flight velocity (0.7 to 2 km/s). The typical test duration for this facility lies in the range of a few tens to a few hundreds of millisecond. Ludwieg tunnel has been extensively used for boundary layer transition \citep{juliano_2008}, skin friction measurement \citep{schulein_2004}, shock boundary layer interaction \citep{neet_2020}, flow over external bodies \citep{labuda_2020}, and so on. As an advancement, high enthalpy flows can also be generated in this facility with the implementation of heated driver gas \citep{segal_2011} and using a free-piston \citep{chung_2018}.

In backdrop of non-availability of enough information on protrusion characteristics at different $Re_D$ and the dominant modes responsible for global unsteadiness, the current experimental campaign aim towards addressing these issues through experimentation in the newly-built Hypersonic Ludwieg Tunnel (HLT). Current investigation focuses on assessing the instabilities associated with sharp protrusions of different $[L/D]$ at various $Re_D$ conditions at $M_\infty = 6$. Attempts are made to understand the source of unsteadiness and quantifying the fluctuations spatiotemporally through high-speed flow visualizations and unsteady pressure responses. The key objectives for the present study include the following:

\begin{itemize}
    \item To design, develop, and calibrate the hypersonic Ludwieg tunnel to study the leading-edge separation problem at $M_\infty=6$ for a wide range of Reynolds numbers based on the freestream conditions and the base body diameter ($Re_D$).
    \item To experimentally generate and vary the leading-edge separation intensity in hypersonic flow using axial protrusions of different lengths ($L/D$) and diameters ($d/D$) from a cylindrical forebody having a base diameter of $D$. 
    \item To estimate the nature of unsteadiness pertaining to the shock and shear layer oscillations through $x-t$ and $x-f$ diagrams constructed from the high-speed schlieren images.
    \item Quantification of the pressure loading ($\kappa$), fluctuation intensity ($\zeta$) and unsteady spectra for protrusion's different  geometrical parameters upon exposure to flow having different $Re_D$.
    \item To identify the dominant spatiotemproal modes responsible for the global unsteadiness through the Proper Orthogonal Decomposition (POD) of the high speed planar laser Rayleigh scattering (PLRS) images.
\end{itemize}

The rest of the manuscript is arranged as follows. Details about the experimental methodology is given in Sec. \ref{sec:exp_meth}. Uncertainty in the measurement techniques are quantified in Sec. \ref{sec:uncertainty}. Vital results from the qualitative and quantitative measurements along with the analysis are provided in Sec. \ref{sec:res_disc}. In Sec. \ref{sec:conclusions}, some of the major conclusions from the present analysis are listed.

\section{Experimental methodology} \label{sec:exp_meth}

\subsection{Facility description}

\begin{figure*}
  \centerline{\includegraphics[width=\textwidth]{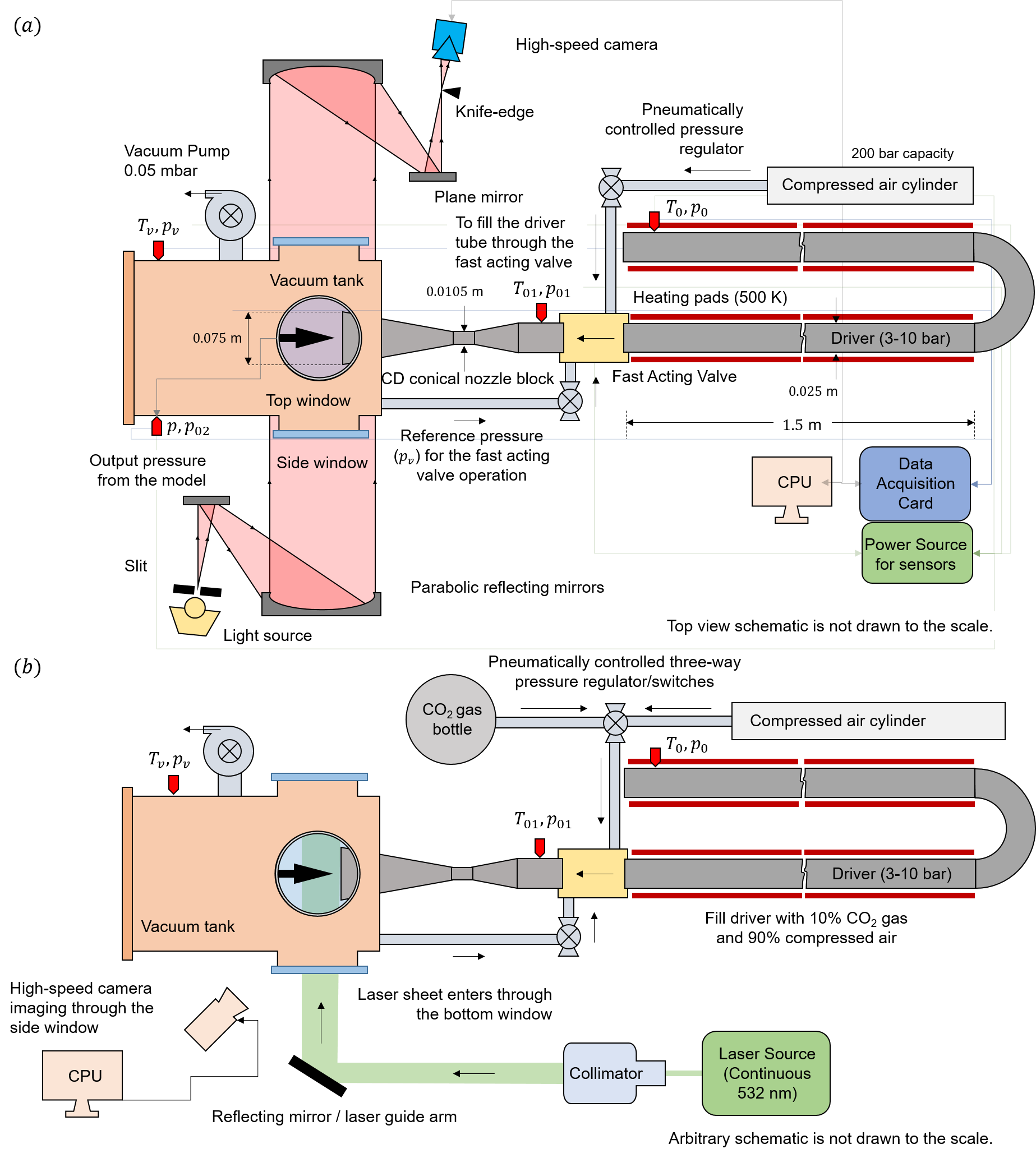}}
  \caption{Typical schematics showing the basic components of the Hypersonic Ludwieg Tunnel (HLT) at Technion which includes driver tube, fast-acting valve, CD nozzle, and vacuum tank. (a) The schematic indicates the arrangement adopted for schlieren imaging and generic pressure measurements. (b) The schematic shows the set-up of planar laser Rayleigh scattering (PLRS) using carbon dioxide as a passive scalar.}
\label{fig:facility}
\end{figure*}

An impulse facility based on the Ludwieg tube operation is proposed to generate hypersonic flow for a short duration \citep{Russell_1973,Li_2022}. A typical layout of the experimental facility along with the optical diagnostics is shown in Figure \ref{fig:facility}. The Ludwieg tube is of 3 m in length and 22 mm in diameter (inner). The tube is kept in the `U'-tube configuration to save some floor space \citep{Schrijer_2010}. The tube is wrapped by heater pads which can maintain a surface temperature up to 500 K. At one end of the tube, a commercial fast-acting valve (ISTA\textsuperscript{\textregistered} KB-20-10) with a valve opening time of $<1$ ms is connected. The tube can be pressurized to a desired fill pressure of $p_0$ between 3 to 10 bar by using the ports in the fast-acting valve itself. The fast-acting valve is operated by computer controlled pneumatic actuators and pressure regulators. There is always a total pressure drop of about $20\%$ in the $p_0$ during the fast-acting valve operation. Hence, a small tube is connected after the valve to measure the actual flow stagnation properties (total pressure and temperature, $p_{01}$ and $T_{01}$). A convergent-divergent (C-D) nozzle is attached next, whose interior is mirror-polished to have a smooth wall surface with a roughness radius of about 0.1 $\mu$m. The nozzle is designed to produce an exit Mach number of $M_d=6.0$. However, after manufacturing, the nozzle's divergent section is measured to have a semi-cone angle of 7$^\circ$ with a throat and exit diameter of 10.4 mm and 76.2 mm, resulting with an actual exit Mach number of $M_e=6.01$. Throughout the manuscript, for the ease of flow representation, the design Mach number of $M_\infty=6.0$ is used. The nozzle expands freely into a closed chamber of volume 0.1 m$^3$ and 0.5 m length. The chamber also possesses four hermetically sealed optical windows (each of 100 mm in diameter) closer to the nozzle exit section and wiring ports for instrumentation downstream. The chamber can be kept to a vacuum level of 0.05 mbar using a rotary vacuum pump (Busch\textsuperscript{\textregistered} RH0021B-Zebra) while monitoring the vacuum pressure level ($p_v$). At the nozzle exit, the freestream Reynolds number ($Re/$m$=u_\infty/\nu_\infty$) is changed by varying the $p_0$ in the Ludwieg tube. A wide range of $Re/m$ between $4.75\times 10^6$ to $1.8\times 10^7$ is achieved while changing $p_0$ between 3 to 10 bar with dry air as the test gas. More details on the freestream conditions from the isentropic relations for $M_\infty=6$ and air as the ideal gas ($\gamma=1.4$) are given in Table \ref{tab:flow_cond}. A computer controlled autonomous environment established using NI (National Instruments) based multi-functional I/O data card observes the necessary pressure ratio between the vacuum tank and the Ludwieg tube to trigger the fast-acting valve, thereby a flow through the nozzle.

\subsection{Freestream calibration}

\begin{figure*}
  \centerline{\includegraphics[width=0.7\textwidth]{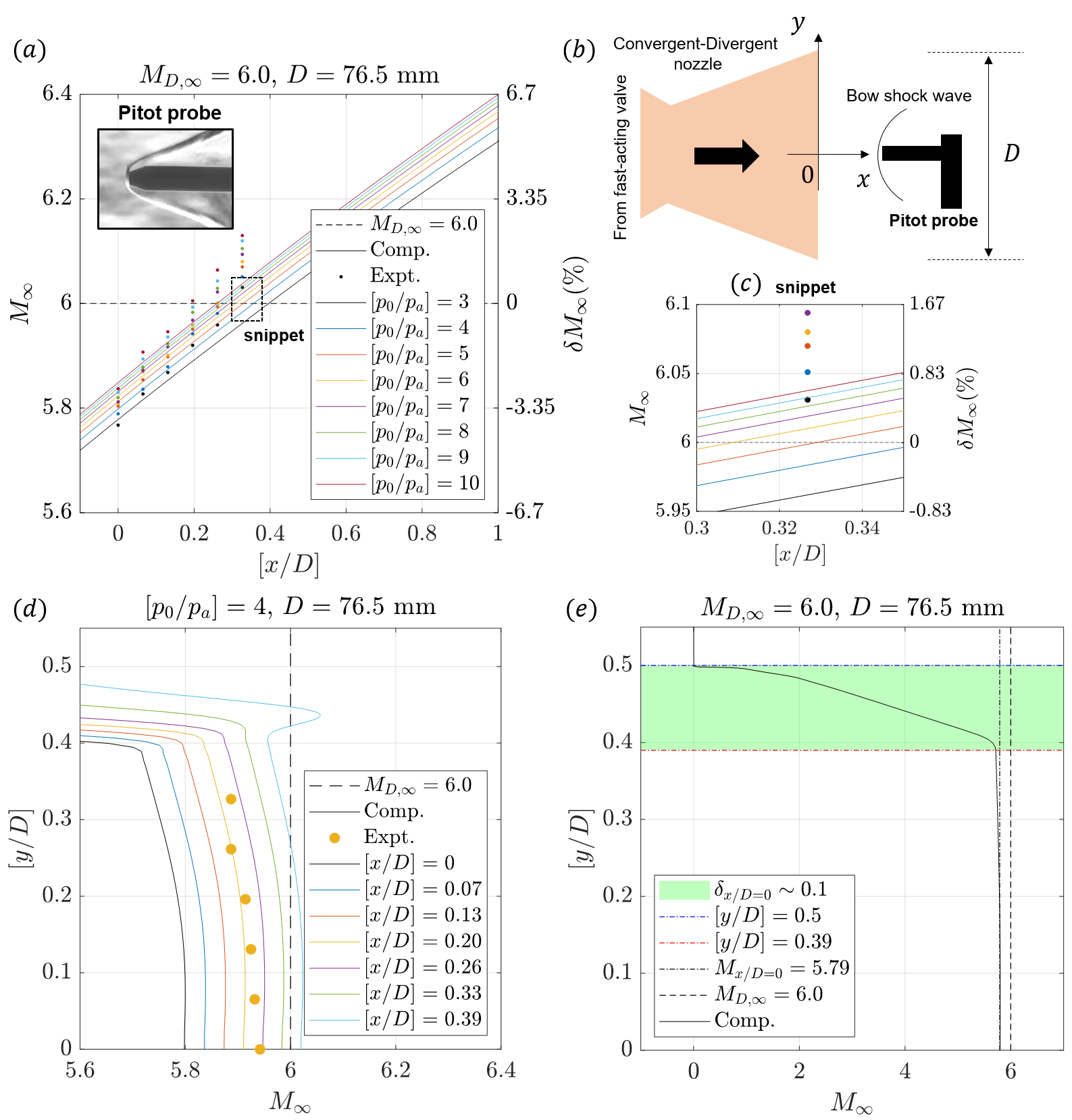}}
  \caption{Plots showing the test-section Mach number calibration. (a) Experimental and computational streamwise Mach number variation at $[y/D]=0$ for different operating conditions with an inset showing the schlieren image around the utilized pitot probe. (b) A schematic showing the pitot probe's position during the data acquisition. (c) A zoomed-in snippet of the streamwise Mach number variation to highlight the deviation between the experiments and computations. (d) Experimental (only at $[x/D]=0.2$) and computational transverse Mach number variation for a particular operating condition ($p_0/p_a=4$). (e) A typical velocity profile at the nozzle exit ($x/D=0$) obtained by solving the Reynolds-Averaged Navier-Stokes equations numerically to highlight the incoming boundary/shear layer thickness ($\delta$).}
\label{fig:calibration}
\end{figure*}

The facility produces a severely under-expanded free jet which provides an open-test section to study the desired models. The test-time of the free-jet lasts only for a brief amount of time after the fast-acting valve's engagement in which the expansion fan travels all along the Ludwieg tube and comes back. In the meantime, a constant flow condition is established in the stagnation chamber and also in the under-expanded jet core. Owing to the expansion process of the jet, the Mach number in the open test-section vary to a certain extent in both streamwise and transverse direction. The flow non-uniformity in the test-section is quantified before the experiments using unsteady pitot probe (see Figure \ref{fig:calibration}) and Rayleigh-pitot formula. The probe has an unsteady pressure transducer (Meggitt\textsuperscript{\textregistered} 8530CM65-100) of head size 3 mm and it is placed perpendicular to the tapered flat-face forebody. The pitot probe's after body is about 12 mm in diameter (see the inset shown in Figure \ref{fig:calibration}a) and extends to a length of at least 10 times its diameter from the model mounting station to avoid any shock-related unsteadiness. Between successive shots of similar flow conditions, the Mach number is measured (Figure \ref{fig:calibration}a) by changing the pitot probe's streamwise distance (Figure \ref{fig:calibration}b). As the facility is automated, each flow condition is repeated conveniently (see Figure \ref{fig:run_time_graphs}). 

As shown in Figure \ref{fig:calibration}, between a streamwise distance of $0\leq [x/D]\leq 0.4$ (here in this sub-section, $D$ represents the nozzle exit diameter), Mach number changes only by 3\%. Between several operating conditions by changing $p_0$, Mach number varies only by $\sim 1.7 \%$. Parallelly, time-averaged computations are done in an axisymmetric jet nozzle using $k-\omega$ SST (shear-stress transport) turbulence closure model. The exhausting jet's computational domain extends up to $10\times 3$ (length $\times$ height) of the nozzle's exit diameter in the streamwise and transverse direction, respectively. The deviation between the experiments and computations are found to be in the order of $\sim 1\%$ in the streamwise measurements (Figure \ref{fig:calibration}c). Transverse pitot-probe measurements are done at a fixed streamwise distance of $[x/D]=0.2$ and the associated Mach number variations from both the experimental and computational analysis are given in Figure \ref{fig:calibration}d. Transverse Mach number variations for the $60\%$ of the jet core are contained within $1\%$. Hence, an effective model diameter of less than 45 mm can be tested. Having a larger size model suffers from non-uniform Mach number and shock interactions due to the presence of the jet shear layer. From the computations (Figure \ref{fig:calibration}e), the exhausting under-expanded jet is measured to have a shear layer thickness of $[\delta/D] \sim 0.1$ at the nozzle exit ($x/D=0$). 

\begin{table*}
\begin{ruledtabular}
\begin{tabular} {l c c c c c c c c}
\textbf{Parameters }& \multicolumn{8}{c}{\textbf{Values}}\\
\hline
Total pressure in the driver supplied ($p_{0} \times 10^5$, Pa) & 3 & 4 & 5 & 6 & 7 & 8 & 9 & 10\\
Total pressure realized in the flow ($p_{01} \times 10^5$, Pa) & 2.31 & 3.25 & 4.14 & 4.80 & 5.70 & 6.48 & 7.37 & 8.13\\
Freestream pressure ($p_{\infty}$, Pa) & 152.64 & 205.84 & 262.21 & 304.01 & 361.02 & 410.42 & 466.79 & 514.92\\
Freestream density ($\rho_{\infty}$, kg/m$^3$) & 0.015 & 0.02 & 0.025 & 0.03 & 0.034 & 0.039 & 0.044 & 0.049\\
Freestream kinematic viscosity ($\nu_{\infty}\times 10^{-5}$, m$^2$/s) & 14.7 & 10.9 & 8.6 & 7.4 & 6.2 & 5.5 & 4.8 & 4.4 \\
Freestream Reynolds number ($Re_D \times 10^5$, $D = 35$ mm) & 1.66 & 2.34 & 2.98 & 3.45 & 4.09 & 4.66 & 5.29 & 5.85\\
Equivalent altitude ($h$, km) & 25.2 & 23.4 & 21.6 & 20.6 & 19.6 & 18.8 & 18.0 & 17.2\\
\hline
Total enthalpy in the driver ($h_{01}$, MJ/kg) & \multicolumn{8}{c}{0.3}\\
Total temperature in the driver ($T_{01}$, K) & \multicolumn{8}{c}{300}\\
Freestream temperature ($T_{\infty}$, K) & \multicolumn{8}{c}{36.58} \\
Freestream velocity ($u_{\infty}$, m/s) & \multicolumn{8}{c}{727.46} \\
Freestream design Mach number ($M_{\infty}$) & \multicolumn{8}{c}{6.0} \\
\end{tabular}
  \caption{Freestream flow conditions for the cases in discussion and the corresponding non-dimensionalized parameters achieved in the test-section. The freestream values are calculated based on the isentropic relations assuming ideal gas flow (air, $\gamma=1.4$) for the considered design $M_\infty$. The only parameter that is changed between the experiments is $p_0$ at a constant $T_{01}$, thereby the freestream $Re_D$ can be changed. $Re_D$ is defined based on the model's base body diameter $D=35$ mm. To non-dimensionalize pressure and time, an atmospheric pressure of $p_a=1 \times 10^5$ Pa and a reference time of $t_0=1$ ms are used wherever needed.}
  \label{tab:flow_cond}
  \end{ruledtabular}
\end{table*}

\subsection{Flow quality and duration}

\begin{figure*}
  \centerline{\includegraphics[width=0.8\textwidth]{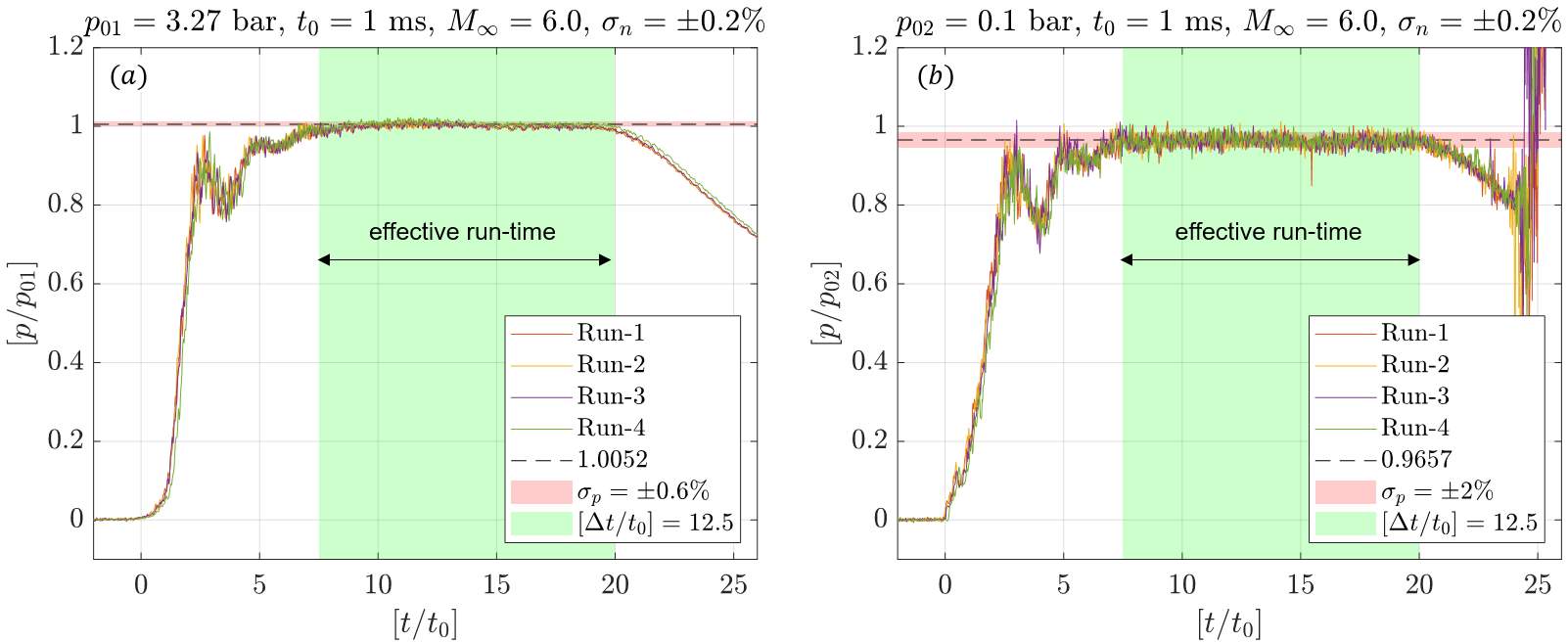}}
  \caption{Typical unfiltered pressure signals showing the repeatability, accuracy, and fluctuations existing (a) in the stagnation chamber and (b) in the freestream, in terms of stagnation ($p_{01}$) and pitot ($p_{02}$) pressure. Time is non-dimensionalized by $t_0=1$ ms. The fluctuations in the ground noise signal ($t/t_0<0$) is $\sigma_n=\pm0.2\%$. The shaded pale-red and light-green regions represent the pressure fluctuation bounds and the obtained effective run-time, respectively. Flow conditions: $[p_0/p_a]=4$, $[p_{01}/p_a]=3.27$, $Re_D=2.34 \times  10^5$, and $M_\infty=6$.}
\label{fig:run_time_graphs}
\end{figure*}

Quantifying the freestream fluctuations and effective run-time are essential before the actual experiments. In Figure \ref{fig:run_time_graphs}, stagnation pressure ($p_{01}$) and freestream pitot pressure ($p_{02}$) variation (see Figure \ref{fig:facility}a for $p_{01}$ and $p_{02}$ measurement location) between the successive runs over a cycle of the expansion process in the Ludwieg tube for a particular flow condition ($[p_0/p_a]=4$, $[p_{01}/p_a]=3.27$, $Re_D=2.34 \times  10^5$) are shown. The signals are unfiltered with a ground noise fluctuation intensity corresponding to $\sigma_n=\pm 0.2\%$. A constant $p_{01}$ is experienced over an effective run-time of 12.5 ms. Theoretically, based on the Ludwieg tube length ($l$) and the sound speed at the local temperature ($a_0$), the expected run-time is calculated as $t_e=[2l/a_0]=[2\times 3/340]\sim 17.7$ ms. However, the transient process due to the valve opening and flow settling in the stagnation chamber consume a part of the flow time. The signal in Figure \ref{fig:run_time_graphs}a is also found to be repeatable between each runs with the standard deviation of $\sigma_p=\pm 0.6\%$ achieved during the effective run-time. Similar events are also observed in the $p_{02}$ measurements at the freestream (Figure \ref{fig:run_time_graphs}b), except for the fact that the fluctuation intensity is amplified to $\sigma_p=\pm 2\%$. One of the primary reasons for the increment is the boundary layer tripping inside the C-D nozzle from a laminar to a turbulent state, thereby introducing freestream fluctuations through weak waves emanation into the core flow \citep{Schneider_1992,juliano_2008}.

\subsection{Model description and instrumentation}

\begin{figure*}
  \centerline{\includegraphics[width=0.85\textwidth]{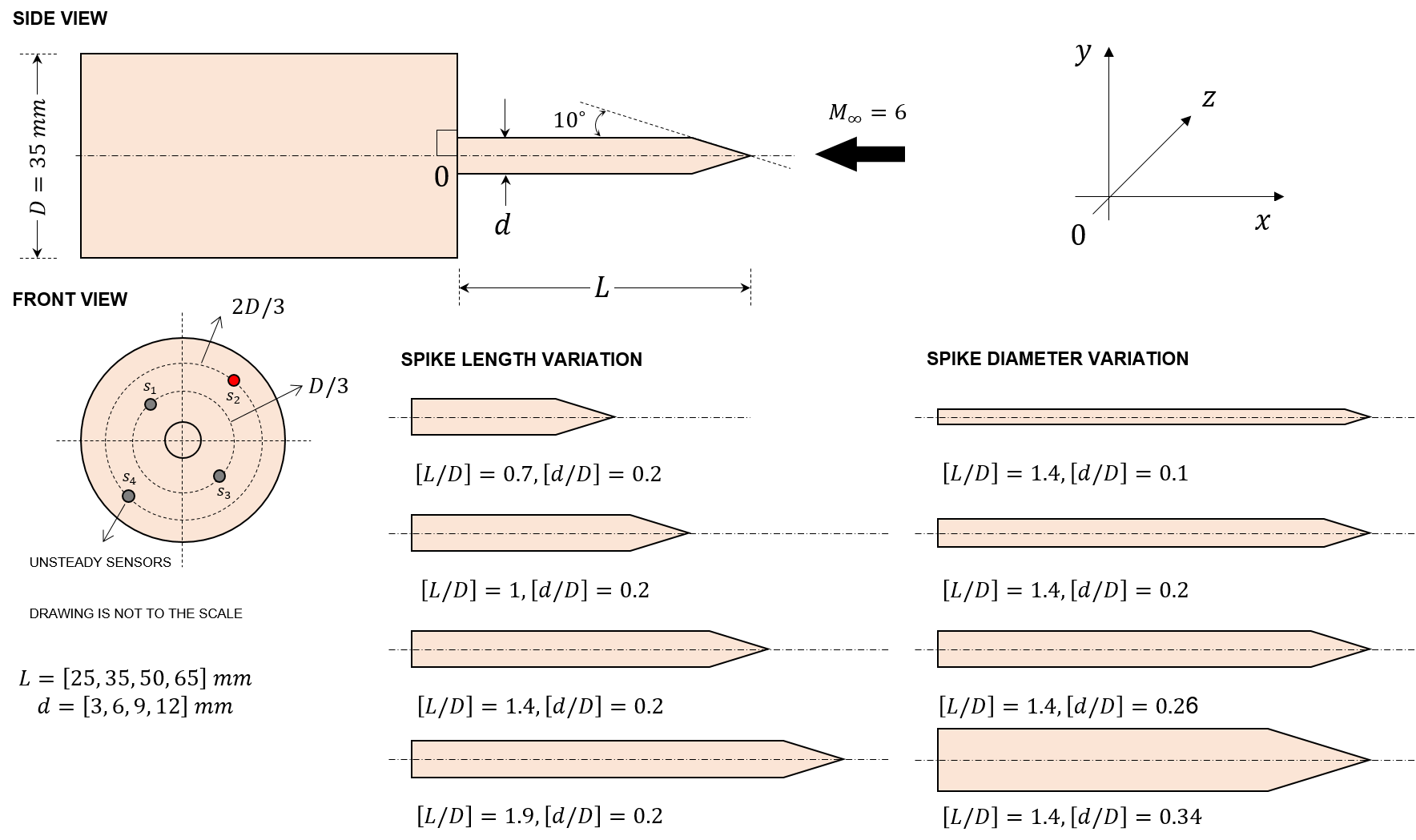}}
  \caption{Schematic showing the utilized model with a flat-face cylindrical forebody having a base body diameter ($D$) and the axial protrusions of different length ($L$) and diameter ($d$). The placement of unsteady pressure transducers on the flat-face cylinder is given where the output of the highlighted sensor (red-colored) is only considered for further discussions.}
\label{fig:model_description}
\end{figure*}

Leading-edge separation is induced by using a flat-face circular cylinder with an axial protrusion as shown in Figure \ref{fig:model_description}. For the experimentation, a flat-face circular cylinder having a base body diameter of $D=35$ mm is made out of perspex. On the forebody's flat-face, drilling is done at two different diameters ($2D/3$, and $D/3$) to mount the unsteady pressure transducers. The protrusion is introduced in the form of a circular cylinder with a sharp tip at one end, whose length ($L$) and diameter ($d$) is varied as $[L/D]=[0.7,1,1.4,1.9]$ and $[d/D]=[0.1,0.2,0.26,0.34]$ in four steps, respectively. The protrusion is made of stainless-steel material with the sharp-tip having a radius of 0.1 mm and $20^\circ$ cone angle. In general the machined surface possess a roughness of 3 to 5 $\mu$m. The machining tolerance is kept at 25 to 30 $\mu$m. The tip of the model with the protrusion is always kept at 5 mm away from the nozzle exit plane.

\subsection{Measurement Methodology} \label{sec:meas_meth}

\begin{figure*}
  \centerline{\includegraphics[width=0.95\textwidth]{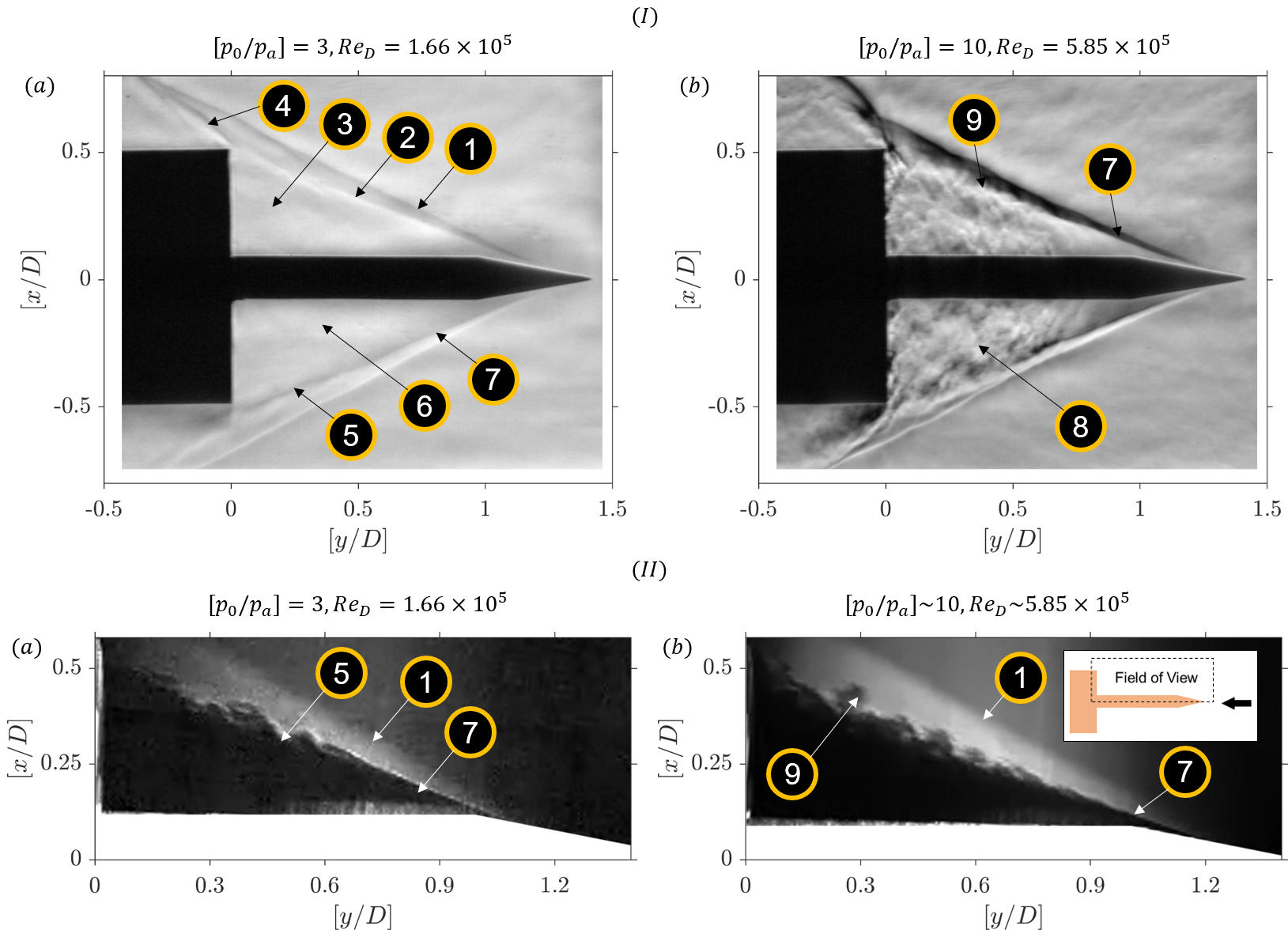}}
  \caption{(I) Normalized instantaneous schlieren image resolving the density gradients ($\parallel \partial \rho/\partial y \parallel$) in the transverse direction and (II) normalized instantaneous planar laser Rayleigh scattering (PLRS) image at an arbitrary time-step during the constant inflow conditions. The considered geometry with an axial protrusion is described by $[L/D]=1.4$, and $[d/D]=0.2$. Each experiment is done at the lowest (a) and the highest (b) $Re_D$ achieved at a constant freestream Mach number of $M_\infty=6.0$. Key flow features in the schlieren and the Rayleigh scattering imaging: 1. leading-edge separation shock; 2. separated shear layer; 3. recirculation bubble; 4. reattachment shock; 5. KH instabilities; 6. moderately quiet recirculation zone; 7. laminar shear layer; 8. noisy recirculation zone; 9. turbulent shear layer.}
\label{fig:hr_inst_sch_plrs_image}
\end{figure*}

Time varying pressure data in the Ludwieg tube, and the vacuum tank are acquired using transducers having 1 ms response time (Omega\textsuperscript{\textregistered} PX309-150GV, and Omega\textsuperscript{\textregistered} Vacuum PX409-015VV). Mean and fluctuating pressure in the stagnation chamber is measured with confidence at two diametrically opposite locations using Meggitt\textsuperscript{\textregistered} 8530CM65-100 and PCB\textsuperscript{\textregistered} 113B28 transducers at 50 kHz. The same PCB\textsuperscript{\textregistered} 113B28 transducer is also used on the flat-face forebody model to calculate the pressure loading ($\overline{p}$) and fluctuation intensity ($\sqrt{\overline{p'^2}}$) at 50 kHz. All the sensors require a supply voltage of 5-10 V to excite the sensing head, and the response are registered in millivolt. A suitable signal conditioning box is used to amplify the signals and are recorded in the NI data acquisition cards. A `k-type' thermocouple capable of measuring 70-700 K with 1 ms response time is used to monitor the flow temperature in the Ludwieg tube, stagnation chamber, and vacuum tank.

A `Z-type' schlieren imaging setup \citep{Settles2001} is used to resolve the density gradients in the transverse direction ($\partial \rho/\partial y$) by placing the knife-edge at the focal point in the horizontal direction. A typical schlieren imaging setup is shown in Figure \ref{fig:facility}-a. The plane and parabolic concave (focal length of 1.5 m) mirrors used in the schlieren system are of 150 mm in diameter. For imaging purpose, a high-speed monochromatic Phantom\textsuperscript{\textregistered} v211 camera is used with an external lens arrangement for projection and focusing. Two set of light sources are used in the imaging process. The first light source from Cavitar\textsuperscript{\textregistered} is used to acquire image at high resolution of 1280 $\times$ 800 pixels and a low frame rate of 1 kHz. The time of light exposure is 150 ns and the images are obtained at 0.1 mm/pixel to capture the details of flow structures at an instant. The second one is a continuous light source from Thor Labs\textsuperscript{\textregistered} which is utilized for a low resolution imaging at 256 $\times$ 128 pixels and at a high repetition rate of 50 kHz. The frame exposure is 2 $\mu$s and the imaging resolution is 0.4 mm/pixel to quantify the dynamics of the shock and the separated shear-layer. A typical schlieren image done using type-I imaging parameters is given in Figure \ref{fig:hr_inst_sch_plrs_image}-I. The moderately quiet and noisy recirculation region is distinctly observed in a typical body with a protrusion at two extreme $Re_D$ to highlight the laminar and turbulent separated shear layer's role. Other common features like the separation, and reattachment shocks are also shown. 

Measurements in a particular streamwise plane ($z/D=0$) is possible using planar laser Rayleigh scattering or PLRS \citep{Do_2010,Zhang_2016}. Carbon dioxide gas is mixed with the compressed air in the Ludwieg tube by 10\% of the desired fill pressure ($p_0$). After the fast-acting valve actuation, the gas expands in the C-D nozzle causing the static temperature to drop, significantly (closer to 36.6 K). Gaseous carbon dioxide sublimes to solid phase by forming ice particles ($<0.5$ $\mu$m). The particles' size are within the incident laser light source's wavelength, thereby scattering the light in the Rayleigh regime (uniform scattering everywhere). The scattering visualizes the vapor cloud's distribution in the freestream and helps in imaging the shock and the separated shear layer. A 10 W continuous 532 nm laser having a 5 mm beam diameter from CNI\textsuperscript{\textregistered} laser (MGL-V-532-10W) is used as the illuminating laser light source. LaVision's\textsuperscript{\textregistered} sheet and collimating optics along with their laser-guiding arm are used to steer the light and convert it into a sheet of uniform width ($< 50$ mm). The scattered light intensity is captured using a Nikon\textsuperscript{\textregistered} 40 mm Micro lens ($f$1/2.8) using the same Phantom\textsuperscript{\textregistered} high-speed camera. The high-resolution imaging is done at 1 kHz, whereas the high-repetition rate imaging is done at 50 kHz. Throughout the manuscript for both the visualizations, the former imaging condition is called as `type-I' imaging, and the later as `type-II' imaging. In general, a series of post-processing routines need to be done on the acquired Rayleigh scattering images to interpret the results, meaningfully. Imaging routines in similar planar laser scattering experiments as described in \cite{Karthick_2017} and \cite{Rao_2020} are proven to be useful in compressible flow field, and they are adopted in the present investigations. A typical PLRS imaging arrangement is explained in Figure \ref{fig:facility}-b. A representative PLRS image obtained using type-II imaging parameters is given in Figure \ref{fig:hr_inst_sch_plrs_image}-II for two extreme $Re_D$ observed in the experimental campaign. The presence of laminar separated shear layer with K-H (Kelvin-Helmholtz) instabilities in the transition zone for the lowest $Re_D$ case and a complete turbulent separated shear layer at the highest $Re_D$ case can also be seen in Figure \ref{fig:hr_inst_sch_plrs_image}-II.

\section{Uncertainty} \label{sec:uncertainty}

Uncertainty in experimental measurements is inevitable. Experiments in hypersonic Ludwieg tunnel is repeated for five times in each conditions. The deviation in the stagnation pressure ($p_{01}$) between successive runs are bounded within $\pm 0.2\%$. Asides, total uncertainty arising from shot repeatability, acquisition methods, data conversion, storage, and other sensitive external factors are considered as suggested by the recommendations in \cite{Coleman2009}. Point measurements including the pressure and temperature acquisitions in the Ludwieg tube and vacuum tank contain 3\% uncertainty. The spatial range across the unsteady pressure logging depends on the sensor diameter, which in this case is about 6 mm. The frequency resolution of the resulting signal is identified identified to 80 Hz. Schlieren imaging poses line-of-sight light integration despite its ability to resolve shocks motion. Spatial resolution is within 0.1 mm and 0.4 mm for type-I and type-II imaging, respectively. Rayleigh scattering being the planar measurement technique, is only limited by the probing laser sheet thickness. In the present experiment, the sheet thickness is calculated to be $\sim$ 1 mm. Spectral resolution for type-II imaging resides within 80 Hz for schlieren imaging and 88 Hz for Rayleigh scattering. Spectra from type-I imaging is seldom used in the manuscript owing to the low sample rate in both the visualization techniques.

\begin{figure*}
  \centerline{\includegraphics[width=\textwidth]{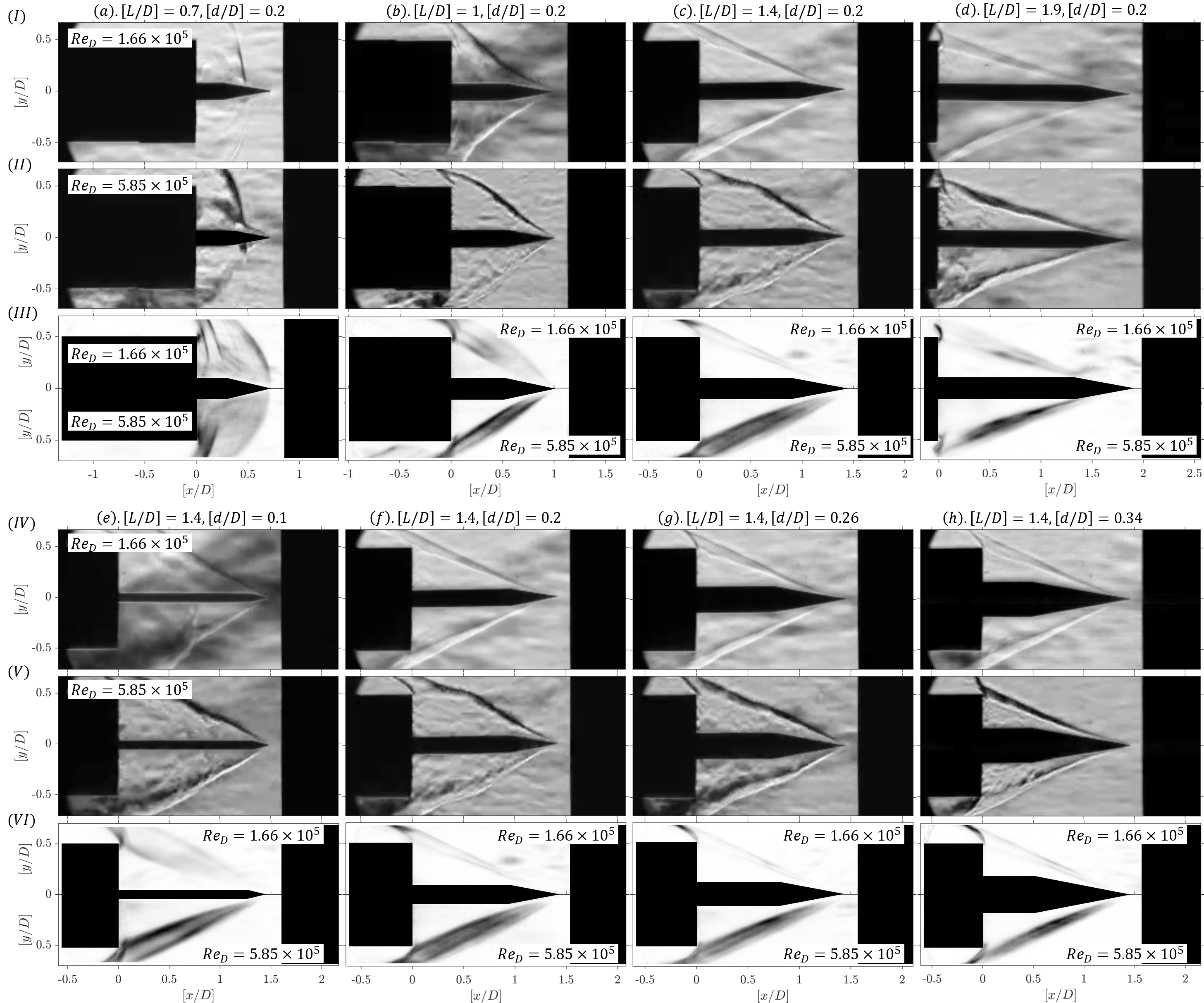}}
  \caption{Normalized instantaneous schlieren image taken at an arbitrary time-stamp during the effective run-time for different cases: (a-d) show $[L/D]=[0.7,1,1.4,1.9]$ variations at a constant $[d/D]=0.2$, and (e-h) show $[d/D]=[0.1,0.2,0.26,0.34]$ variations at a constant $[L/D]=1.4$. Images are also provided for the two extreme $Re_D$ encountered in the experiments: (I, IV) $Re_D=1.66 \times 10^5$, and (II,V) $Re_D=5.85 \times 10^5$. (III, VI) Normalized operator based time-averaged image ($\boldsymbol{R}$) showing the mean characteristics of the fluctuating flow field, where owing to the flow symmetry and brevity, the top and bottom halves are only represented for the cases of $Re_D=1.66 \times 10^5$ and $Re_D=5.85 \times 10^5$ while varying $[L/D]$ and $[d/D]$. See the \href{https://youtu.be/AroYgvT9OCs}{supplementary} for viewing the corresponding video files.}
\label{fig:inst_sch_image_all}
\end{figure*}

\section{Results and discussions} \label{sec:res_disc}
In this section, qualitative and quantitative measurements are provided and the findings are discussed.

\subsection{Qualitative imaging}
In qualitative imaging, two visualization methods are adopted: Schlieren imaging and Planar Laser Rayleigh Scattering.

\subsubsection{Schlieren imaging}
Normalized instantaneous schlieren images are obtained using type-II imaging (see Sec. \ref{sec:meas_meth}) for each variation in $[L/D]$ and $[d/D]$ as shown in Figure \ref{fig:inst_sch_image_all}. Schlieren images resolve the density gradients along the transverse direction by cutting the focal-point using a horizontal knife edge. The density gradient image is subjugated through a mathematical processing routine to bring out the features of the fluctuating flow field with clarity as described by the equations,
\begin{align}
\label{eq:den_trans}
R &= \partial \rho / \partial y ,\\ 
\label{eq:op_avg}
\boldsymbol{R} &= \lVert \overline{R} - R_{rms} \rVert.
\end{align}
In Figure \ref{fig:inst_sch_image_all} I-a and II-a (also see the \href{https://youtu.be/AroYgvT9OCs}{supplementary} video), the normalized instantaneous schlieren images reveal the `pulsation' form of shock oscillation for $[L/D,d/D]=[0.7,0.2]$. The shock travel to-and-fro between the protrusion's tip and the forebody at fixed frequency. The case is observed to be persistent throughout the $Re_D$ cases under discussion. On the other hand, all the remaining cases exhibit `flapping' form of shock oscillation. During flapping, the separated shear layer and shock moves inwards and outwards with separation point as the hinge. Flapping is severe especially at the highest $Re_D$ ($5.85 \times 10^5$). The least $Re_D$ case ($1.66 \times 10^5$) exhibit only a feeble flapping which is even difficult to notice. Visually, for all the $[L/D]$ and $[d/D]$ variations, the images look sharp and have a good contrast for the highest $Re_D$. The presence of larger density in the freestream at the highest $Re_D$ (see Table \ref{tab:flow_cond}) might be the reason for such an image contrast between the cases. From the qualitative analysis on instantaneous images for $[L/D]$ variations (Figure \ref{fig:inst_sch_image_all}-II a-d), flapping is seen for $[L/D]\geq 1$. However, the flapping intensity seems to be reducing with increase in $[L/D]$. Likewise, for $[d/D]$ variations (Figure \ref{fig:inst_sch_image_all}-V e-h), flapping intensity decreases with increments in $[d/D]$.

In Figure \ref{fig:inst_sch_image_all} III and VI, operator based time-averaged images are given to quantitatively draw some conclusions on the intensity of shock-related unsteadiness. The top and bottom half of the image represent the flow field at the lowest and the highest $Re_D$, respectively. The lowest $Re_D$ case ($1.66 \times 10^5$) possess laminar separated shear layer for most of the part with incoming flow also being laminar. Hence, the regions of shade seen at low $Re_D$ case are minimal. It also means that the resulting fluctuations or shock-related unsteadiness are not significantly pronounced. On the other hand, for the highest $Re_D$ case ($5.85 \times 10^5$), the shaded zones are more prominently seen, especially around the separated shear layer and shock. The reason is attributed to the unsteadiness in the separated shock and shear layer motion arising from the laminar to turbulent transition in the separated shear layer. From the high resolution type-I imaging shown in Figure \ref{fig:hr_inst_sch_plrs_image} I-b, the acoustic waves are seen to be populating the recirculation zone which is confined by the forebody, separated shear layer, and protrusion. It is suspected that these acoustic waves travelling between the forebody and the separation point might excite the separated shear layer to flap vigorously. However, asides these images, there is no experimental evidence to further support the postulate. 

For the $[L/D,d/D]=[0.7,0.2]$ case (Figure \ref{fig:inst_sch_image_all} III-a), the presence of a hanging shock in the protrusion's tip and shocks standing-off the forebody surface represent the to-and-fro shock motion in pulsation type of shock oscillations. The image is also in agreement with a similar pulsating flow field observed at supersonic Mach numbers in the experiments of \cite{sahoo_2021}. Asides, the shaded zones vary from thick to thin as we change the $[L/D]$ (Figure \ref{fig:inst_sch_image_all}-III b-d) and $[d/D]$ (Figure \ref{fig:inst_sch_image_all}-VI e-h), indicating the reduction in flapping intensity.

\begin{figure*}
  \centerline{\includegraphics[width=\textwidth]{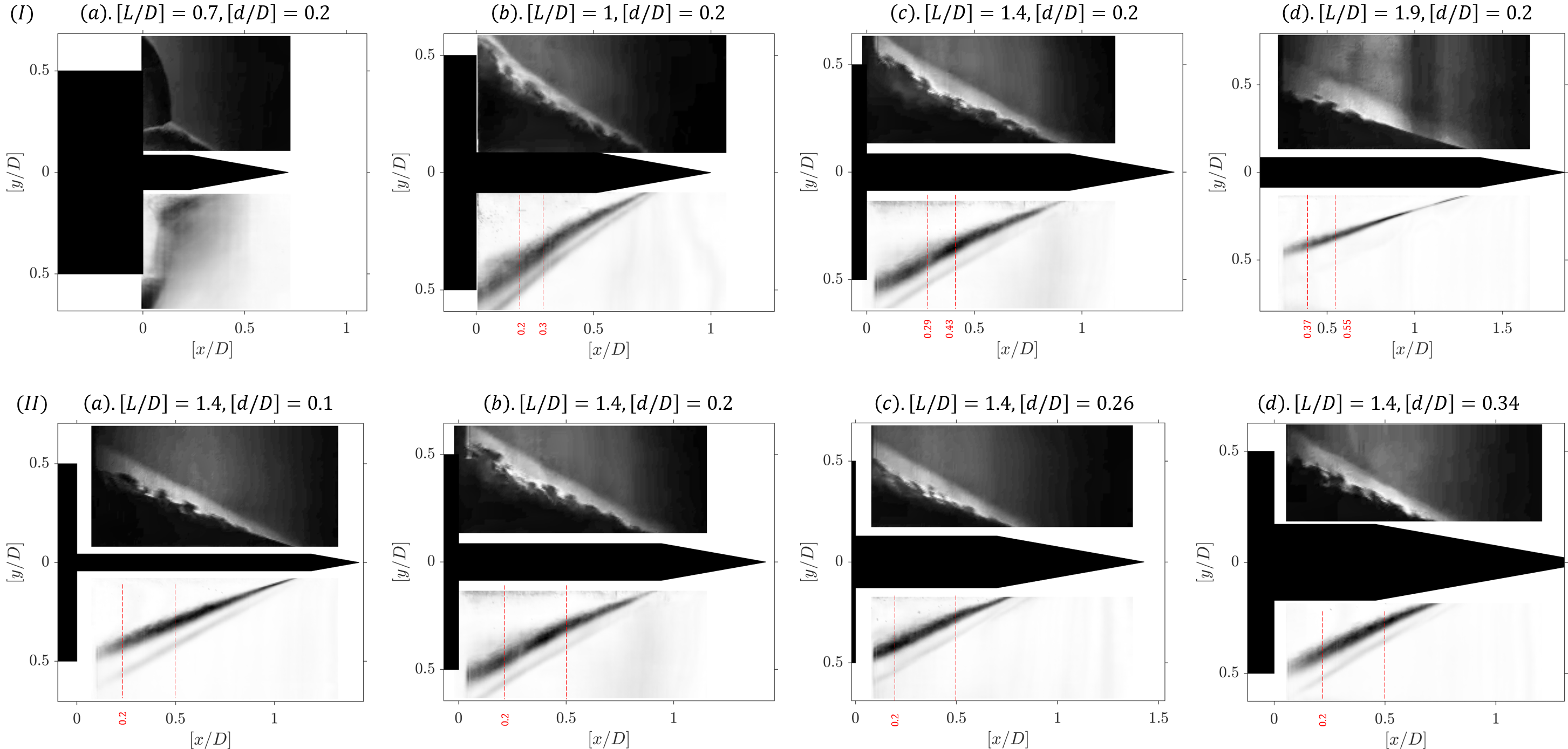}}
  \caption{Normalized instantaneous Rayleigh scattering image taken at an arbitrary time-stamp during the effective run-time for different cases: (I, a-d) show $[L/D]=[0.7,1,1.4,1.9]$ variations at a constant $[d/D]=0.2$, and (II, a-d) show $[d/D]=[0.1,0.2,0.26,0.34]$ variations at a constant $[L/D]=1.4$. Images are only provided for the highest $Re_D$ encountered in the experiments: $Re_D=5.85 \times 10^5$. In each image, the bottom half represents the normalized operator based time-averaged image ($\boldsymbol{R}$) shown in an inverted color, displaying the mean characteristics of the fluctuating flow field. The two dotted vertical lines in each image denote the location at which the intensity profiles are investigated to deduce the separated shock and shear layer properties as described in Sec. \ref{sec:shr_lyr_prop}. See the \href{https://youtu.be/JVMmbYXe0Pg}{supplementary} for viewing the corresponding video files.}
\label{fig:plrs_mean_inst}
\end{figure*}

\subsubsection{Planar laser Rayleigh scattering}

Normalized instantaneous Rayleigh scattering images (see also the \href{https://youtu.be/JVMmbYXe0Pg}{supplementary} video) obtained using type-II imaging parameters are given in the top half of Figure \ref{fig:plrs_mean_inst} for variations in $[L/D]$ and $[d/D]$ at the highest $Re_D$ of $5.85 \times 10^5$. The bottom half contains the operator based time-averaged image with inverted color map for clarity. From the instantaneous images, for $[L/D,d/D]=[0.7,0.2]$ (Figure \ref{fig:plrs_mean_inst}-Ia), pulsating form of shock unsteadiness is seen. In general, pulsation cycle contains three phases: inflate, collapse, and with-hold as reported in the literature \citep{Feszty2004a}. In the shown time instant, flow field during the collapsing phase is shown. The triple point of the shock system is clearly seen. Ahead, the flow is seeded, however, due to unsteady shock motion, flow is severely recirculating. Hence, the downstream is not filled with any particles. The operator based time-averaged image shows a smudgy patches marking the unsteady shock/shear layer passage in the flow field. The image is consistent with the time-averaged schlieren image reported in the pulsating flow at supersonic speed \citep{sahoo_2021}.

For all other cases of $[L/D]\geq 1$, as told in the earlier section, flapping form of shock unsteadiness is observed with decreasing flapping intensity. While varying the $[L/D] \geq 1$ (Figure \ref{fig:plrs_mean_inst}-I b-d), flow deflection angle ($\theta$) produced by the separated shear layer, and the oblique shock angle ($\beta$) originating about the separation point monotonically decreases. Moreover, the separated shear layer transitions from a laminar to a turbulent state earlier, for shorter $[L/D]$ and vice versa for the longer one. For example in the case of $[L/D, d/D]=[1,0.2]$, the transition point ($x/D \sim 0.8$) is closer to the separation point ($x/D \sim 0.75$). However, in the case of $[L/D, d/D]=[1.9,0.2]$, the transition point ($x/D \sim 0.8$) is located far away from the point of separation ($x/D \sim 1.5$). Similarly, the separated shear layer thickness from the operator based time-averaged image reveal a thick shear layer for the shorter $[L/D]$ and a thin one for the longer case. While varying $[d/D]$, from Figure \ref{fig:plrs_mean_inst}-II a-d and the \href{https://youtu.be/JVMmbYXe0Pg}{supplementary} video, the flapping intensity is seen to be decreasing with increments in $[d/D]$. The values of $\theta$ and $\beta$ are also qualitatively seen to be the same. Similarly, the transition point in the separated shear layer lies close to the separation point for almost all the cases of $[d/D]$. More quantitative details about the shear layer properties are discussed in the upcoming section (see Sec. \ref{sec:shr_lyr_prop}).

\subsection{Quantitative assessments}

\begin{figure*}
  \centerline{\includegraphics[width=\textwidth]{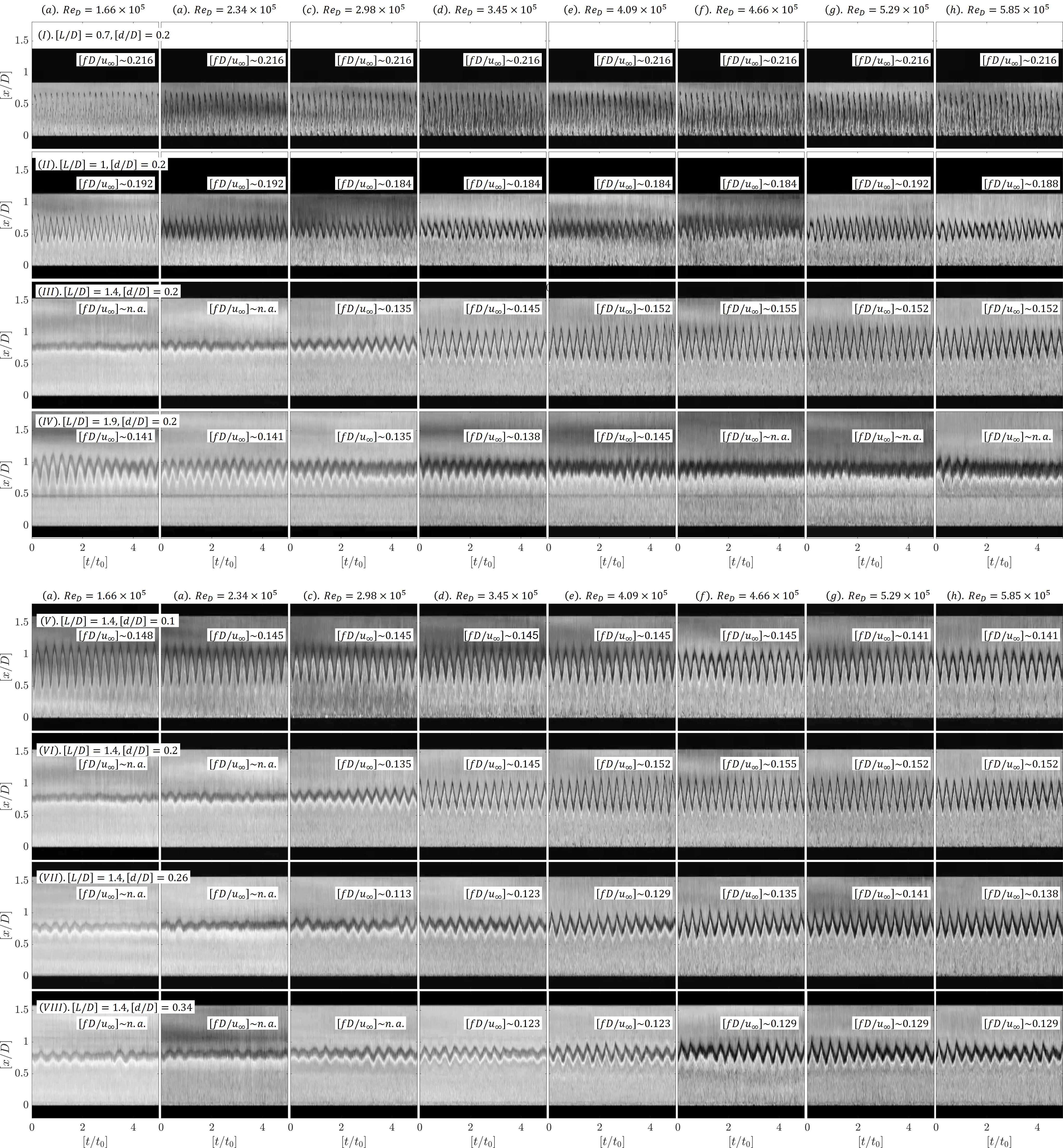}}
  \caption{A series of $x-t$ plots showing the shock and separated shear layer motion over a particular region of effective run-time for a wide range of variations in $[L/D]$, $[d/D]$, and $Re_D$. The $x-t$ plot is constructed by taking the light intensity profile at $[y/D]=0.25$ on the instantaneous schlieren images that are obtained using type-II imaging (see Figure \ref{fig:inst_sch_image_all} I-II and IV-V). In each $x-t$ plot, the dominant non-dimensionalized frequency due to the discrete shock/shear layer oscillation is given at the top-right from the respective $x-f$ diagram (fast Fourier transform analysis of the $x-t$ plot).}
\label{fig:xt_plot_all}
\end{figure*}

\subsubsection{$x-t$ and $x-f$ analysis}
\begin{figure*}
  \centerline{\includegraphics[width=\textwidth]{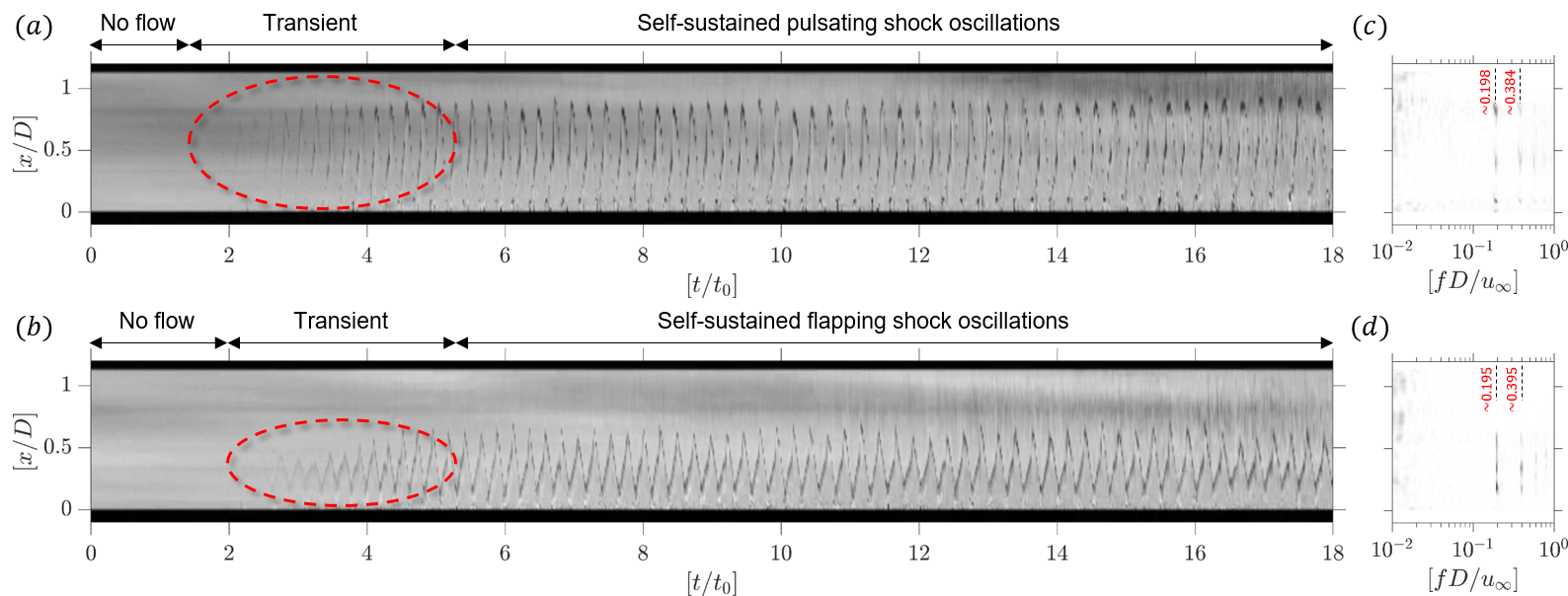}}
  \caption{Typical (a-b) $x-t$ and (c-d) $x-f$ plots shown for the same geometrical case of $[L/D]=1$, and $[d/D]=0.2$ from a high-speed schlieren imaging. The $x-t$ plot is constructed by stacking the light intensity profile at $[y/D]=0.25$ from the beginning of the tunnel operation. The fast Fourier transform analysis shown in the $x-f$ plot (c-d) is performed only for the duration of the self-sustained shock oscillation as marked in (a-b). The dotted ellipse in (a-b) marks the transient flow patterns seen during a typical run where the operating conditions are held constant. Flow conditions: $[p_0/p_a]=3$, $[p_{01}/p_a]=2.31$, $Re_D=1.66 \times  10^5$, and $M_\infty=6$. }
\label{fig:dual_mode_unsteadiness}
\end{figure*}

\begin{figure*}
  \centerline{\includegraphics[width=\textwidth]{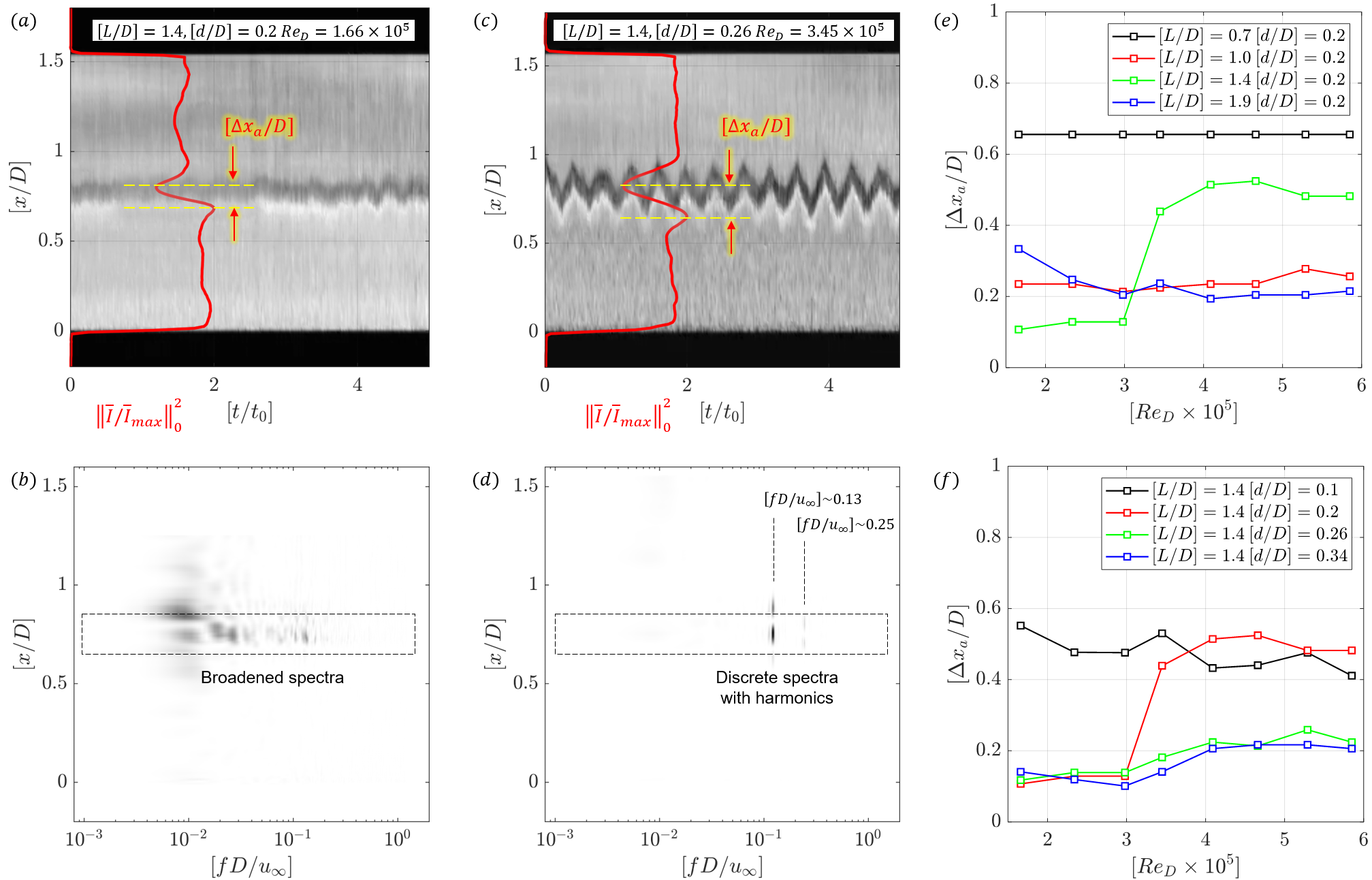}}
  \caption{(a,c) Typical $x-t$ and (b,d) $x-f$ plots revealing the presence of broadened and discrete spectra from the mild and severe shock/shear layer oscillations obtained from two specific cases. In the $x-t$ plots, the normalized time-averaged light intensity profile along the $x$-direction is given in red-color. The local extrema in the profile marks the extent of shock/shear layer oscillations ($\Delta x_a/D$). (e-f) Plots showing the variations of $\Delta x_a/D$ for changes in $[L/D]$ and $[d/D]$ across different $Re_D$.}
\label{fig:xt_xf_del_xa}
\end{figure*}

The extent and nature of unsteady shock oscillations along with the driving frequency are understood quantitatively using $x-t$ and $x-f$ plots. The variations in the aforementioned quantities are studied for different $[L/D]$, $[d/D]$, and $Re_D$ cases. Firstly, a series of $x-t$ plots is constructed by stacking the light intensity variation from an instantaneous schlieren image at a particular line profile of $[y/D]=0.25$ over the effective run-time. Such a procedure creates an image containing the motion of the separation shock foot and shear layer with a good contrast. Performing a fast Fourier transform on the $x-t$ plots lead to the formulation of $x-f$ plots that reveal spectral contents of those time varying events. A series of $x-t$ plots is given in Figure \ref{fig:xt_plot_all} for a wide range of variations in $[L/D]$, $[d/D]$, and $Re_D$. An example of both $x-t$ and $x-f$ plots is only given for two particular cases in Figure \ref{fig:xt_xf_del_xa} a-d for brevity. For some of the cases, the $x-t$ plot reveal the wave pattern being broadband with a small amplitude like in the case of $[L/D,d/D]=[1.4,0.2]$ at $Re_D=1.66 \times 10^5$ (Figure \ref{fig:xt_xf_del_xa}a). The corresponding $x-f$ plot also confirms the presence of broadened spectra (Figure \ref{fig:xt_xf_del_xa}b). The wave patterns sometimes look like saw-tooth as in the $x-t$ plot (Figure \ref{fig:xt_xf_del_xa}c) of $[L/D,d/D,Re_D]=[1.4,0.26,3.45 \times 10^5]$ whose $x-f$ plot (Figure \ref{fig:xt_xf_del_xa}d) reveals the presence of a discrete dominant non-dimensional frequency.

From the initial assessment of the $x-t$ plots (Figure \ref{fig:xt_plot_all}-I a-h), the case of $[L/D,d/D]=[0.7,0.2]$ (shortest) produces pulsating mode of unsteadiness as observed from the qualitative schlieren (Figure \ref{fig:inst_sch_image_all} III-a) and the Rayleigh scattering (Figure \ref{fig:plrs_mean_inst}-Ia). The pulsating shock leaves a high frequency sinuous wave pattern as a trace. The extent of shock oscillation or the amplitude of the sine wave is bounded between the entire region of $0 \leq [x/D] \leq 0.7$. From the $x-f$ analysis, the dominant pulsation frequency is computed to be $[fD/u_\infty] \sim 0.22$ and it remains constant with $Re_D$. The pulsating frequency is also found to be in agreement with the empirical relation proposed by \cite{Kenworthy1978} as given in the equation below.
\begin{align}
    \left[\frac{fD}{u_\infty}\right] = 0.25 - \left(\frac{L}{D}\right)0.067.
\end{align}
In the empirical relation, the independence of the pulsating frequency with $Re_D$ can also be seen. The non-dimensional experimental frequency is having a deviation of $\pm$ 6.4\% from the empirical relation.

While changing the $[L/D]$ to 1, an interesting flow feature is observed for repeated experiments. The pattern of shock unsteadiness continuously changes from pulsation to flapping or vice versa between every runs irrespective of $Re_D$. Although the upstream and the downstream conditions are kept almost to a constant condition based on the available instrumentation, the varying pattern change is inevitable. In order to understand the trigger for such a change between the two form of shock unsteadiness, $x-t$ and $x-f$ plots are constructed at a specific $Re_D$ of $1.66 \times 10^5$ as shown in Figure \ref{fig:dual_mode_unsteadiness}. The $x-t$ plots are drawn from the beginning of the test till the end of the effective run-time. However, the $x-f$ plots are drawn based on the signal obtained only during the effective run-time, where a self-sustained shock oscillation is observed. While reviewing the $x-t$ plots, the wave patterns seen during the transient portion of the run-time are observed to be different. A self-sustained pulsating shock oscillation is seen if the fluctuations in the transient portion is severe. Whereas, a self-sustained flapping shock oscillation is seen for a mild fluctuation in the transient zone. The reason for the observation of different fluctuation types during the transient period can be attributed to the flow quality changes between the runs owing to the slightest non-repeatable movement of the fast-acting valve. The $x-f$ plot for pulsation and flapping, however, reveal a same non-dimensional frequency and harmonics at $[fD/u_\infty] \sim 0.2$. Thus, it can be inferred that the initial disturbances seen in the transient zone has no effect on the non-dimensional frequency pertaining to shock oscillation. Asides, it has to be noted that the other cases of $[L/D]$ and $[d/D]$ indeed offer a repeatable unsteady signature for a wide range of $Re_D$. They are not affected by the smallest imperfections in the fast-acting valve operation. It is just this particular case, which is so sensitive to the upstream disturbances. For the rest of the discussion pertaining to this particular case, we consider only the self-sustained flapping motion, as the authors are more inclined to investigate the flapping behavior.

For cases of $[L/D] \geq 1$, the shock signature looks more like a saw-tooth wave with a considerably lower frequency than the pulsating shock's case, especially for moderate $Re_D$. In general for $[L/D,d/D]=[1,0.2]$, the dominant non-dimensionalized frequency varies between $0.18 \leq [fD/u_\infty] \leq 0.19$ in an arbitrary manner with $Re_D$. While increasing the $[L/D]$, as in the case of $[L/D,d/D] = [1.4,0.2]$, no noticeable fluctuations are seen in the shock foot for $1.66 \times 10^5 \leq Re_D \leq 2.34 \times 10^5$. However, a transition in shock motion (from a broad low-amplitude signal to a discrete saw-tooth signal of considerable amplitude) is observed from $Re_D \geq 2.98 \times 10^5$ with the dominant non-dimensionalized frequency varying between $0.13 \leq [fD/u_\infty] \leq 0.15 $. The formation of the laminar separated shear layer at lower $Re_D$ is suspected to be responsible for less intense broadband signal. For the case of $[L/D,d/D]=[1.9,0.2]$, the shock motion transitions from a discrete saw-tooth wave to more of a broadband wave with significant amplitude, especially after $Re_D \geq 4.66 \times 10^5$. Owing to the length of the protrusion, the separated shear layer undergoes transition from a laminar to a turbulent state at high $Re_D$. It leads to the observation of a broadened signal with a significant amplitude in the $x-t$ plot. During the discrete flapping motion, the dominant non-dimensionalized frequency varies between $0.13 \leq [fD/u_\infty] \leq 0.14$. 

While varying $[d/D]$, as in the case of $[L/D, d/D]=[1.4,0.1]$, a distinct saw-tooth wave pattern of shock oscillation is observed irrespective of $Re_D$. However, as the $[d/D]$ increases, particularly for $[L/D, d/D]=[1.4,0.2]$, initial set of $Re_D$ cases ($1.66 \times 10^5 \leq Re_D \leq 2.34 \times 10^5$) exhibit broadened low-amplitude oscillations and later transforms in to a discrete saw-tooth wave at higher $Re_D$, as explained in the previous paragraph. Pushing the $[d/D]$ further, only delays the occurrence of discrete flapping to higher $Re_D$. For example, in the case of $[L/D, d/D]=[1.4,0.26]$, discrete waveform is seen for $Re_D \geq 2.98 \times 10^5$ with $0.12 \leq [fD/u_\infty] \leq 0.14$. Moreover, for $[L/D, d/D]=[1.4,0.34]$, the transition occurs at $Re_D \geq 3.45 \times 10^5$ with $0.12 \leq [fD/u_\infty] \leq 0.13$.

\subsubsection{Extent of shock/shear layer oscillation} \label{sssec:shk_shr_osci}

Asides the frequency of the wave pattern left behind by the shock motion, the extent of the shock motion along the $x$-direction is also a good indicator to understand the intensity of flapping. It has to be emphasized that the flapping motion is traced both from the shock and the separated shear layer as they move in harmony and stay close to each other at high freestream Mach numbers. In order to extract the extent of shock motion, the $x-t$ diagram is subjected through a series of image processing routines. Intensity variation along the $x$-direction is averaged over time. The values are normalized between 0 to 2 and plotted in red-color as shown in Figure \ref{fig:xt_xf_del_xa} a and c. In the region of interest where the shock motion is relevant, local extrema is seen in the time-averaged intensity profile. The distance between the local two peaks ($\Delta x_a/D$) is used as a measure to quantify the extent of shock motion. All the $x-t$ plots shown in Figure \ref{fig:xt_plot_all} are subjected to these processing routines and the corresponding $[\Delta x_a/D]$ are calculated. Later, these values are plotted to see the variations with respect to $[L/D]$ and $[d/D]$ over a range of $Re_D$ as shown in Figure \ref{fig:xt_xf_del_xa} e-f.

\begin{figure*}
  \centerline{\includegraphics[width=0.75\textwidth]{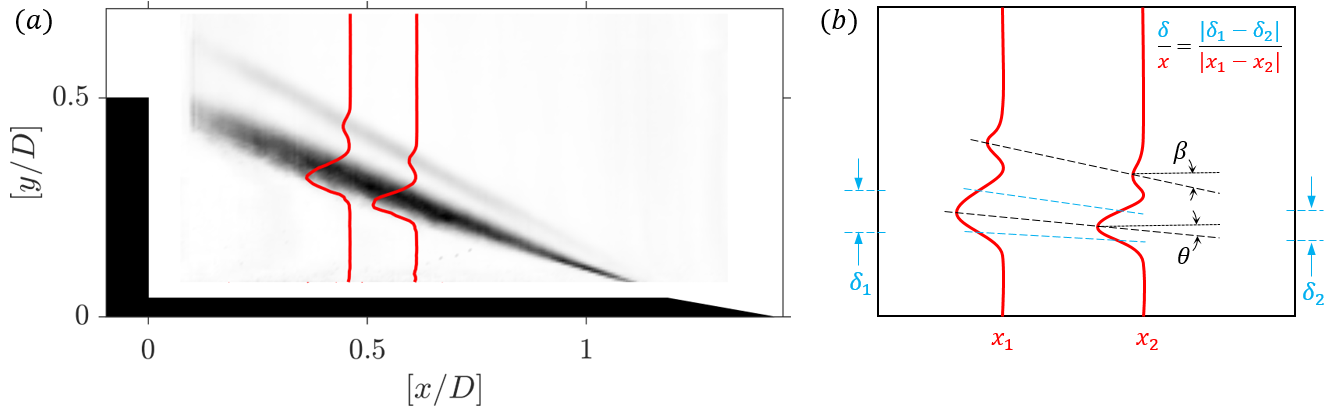}}
  \caption{(a) A typical operator based time-averaged image from the Rayleigh scattering type-II imaging. An arbitrary intensity profile along the $y$-direction at two $x$-locations marks the presence of shock and shear layer. (b) A schematic showing the procedure to extract the apparent shear layer properties including the shear layer angle ($\theta$), shock angle ($\beta$), and shear layer growth rate ($\delta/x$). }
\label{fig:shr_lyr_prop_fig}
\end{figure*}

\begin{table*}
\begin{ruledtabular}
\begin{tabular} {c c c c c c c c c}
\multirow{2}{*}{Cases} & \multicolumn{4}{c}{$[d/D]=0.2$} & \multicolumn{4}{c}{$[L/D]=1.4$}\\ \cmidrule(l){2-9} 
 & $[L/D]=0.7$ & $[L/D]=1$ & $[L/D]=1.4$ & $[L/D]=1.9$ & $[d/D]=0.1$ & $[d/D]=0.2$ & $[d/D]=0.26$ & $[d/D]=0.34$ \\ \cmidrule(r){1-9}
$\theta$ $(^\circ)$   & - & 37.88 & 24.62 & 15.94 & 23.05 & 28.01 & 25.08 & 25.20 \\ 
$\beta$ $(^\circ)$   & - & 41.63 & 30.26 & 24.90 & 28.01 & 29.88 & 30.78 & 30.46 \\ 
$\delta /x$ $(-)$ & - & 0.35  & 0.18  & 0.05  & 0.04  & 0.08  & 0.05  & 0.06 \\ 
\end{tabular}
  \caption{Tabulation of the apparent shear layer properties for a wide range of $[L/D]$ and $[d/D]$ at the largest $Re_D$ of $5.85 \times 10^5$.}
  \label{tab:shr_lyr_prop}
  \end{ruledtabular}
\end{table*}

While monitoring the changes in $[\Delta x_a/D]$ by varying $[L/D]$ and $Re_D$ (see Figure \ref{fig:xt_xf_del_xa}e), a constant value of $[\Delta x_a/D] \sim 0.65$ is observed over the entire range of $Re_D$, particularly for the pulsating case ($[L/D,d/D]=[0.7,0.2]$). While increasing the $[L/D]$ to 1, the value of $[\Delta x_a/D]$ is observed to be considerably lower than the previous case as it remains almost a constant ($[\Delta x_a/D] \sim 0.23$) over different $Re_D$. Further increasing the length to $[L/D]=1.4$, values of $[\Delta x_a/D]$ is even lesser for laminar range of $Re_D$ with $[\Delta x_a/D] \sim 0.1$. However, after the critical $Re_D$ of $2.98 \times 10^5$, the values of $[\Delta x_a/D]$ increases drastically to almost $[\Delta x_a/D] \sim 0.5$ for the rest of the $Re_D$ under consideration. While analysing the respective $x-t$ plots (Figure \ref{fig:xt_plot_all}-III d-h) in the critical $Re_D$, one can see the shock/shear layer motion pattern transforming from a broadened low-amplitude wave to a high-amplitude saw-tooth wave. For the longest $[L/D]$ of 1.9, values of $[\Delta x_a/D]$ is observed to be very high ($[\Delta x_a/D] \sim 0.35$) for a laminar $Re_D$, initially. While changing from lower to higher $Re_D$, $[\Delta x_a/D]$ values are observed to be decreasing and asymptotically becomes a constant at $[\Delta x_a/D] \sim 0.2$. The reason behind the asymptotic trend can once again be explained through the $x-t$ plots. For the longer $[L/D]$ case, the separated shear layer travels to a longer distance from the separation point to the forebody shoulder. While doing so, the separated shear layer undergoes transition from a laminar to a turbulent state, especially at higher $Re_D$. The $x-t$ plots (Figure \ref{fig:xt_plot_all}-IV a-h) are constructed by taking light intensity profiles at $[y/D]=0.25$ which intersects the flow closer to the forebody shoulder. Hence, the light intensity variations contain only the turbulent motion of the structures in the shear layer, which are generally broadband in nature. Hence, the patterns of shock/shear layer motion changing from saw-tooth to broadband at higher $Re_D$ lead to the observation of asymptotic values in $[\Delta x_a/D]$.

While changing the $[d/D]$, variations of $[\Delta x_a/D]$ with $Re_D$ are plotted in Figure \ref{fig:xt_xf_del_xa}f. For the thinnest protrusion ($d/D=0.1$), globally the values of $[\Delta x_a/D]$ decrease as $0.57 \geq [\Delta x_a/D] \geq 0.4$ with $Re_D$. As seen from the respective $x-t$ plots (Figure \ref{fig:xt_plot_all}-V a-h), the decrement in flapping intensity is attributed to the trend seen in $[\Delta x_a/D]$. In the previous paragraph, it is explained how the turbulent separated shear layer produces a low $[\Delta x_a/D]$. In also the present case, the moderate stabilization of flapping at higher $Re_D$ due to the turbulence in the separated shear layer is held responsible for the damping behaviour in $[\Delta x_a/D]$. The next $[d/D]$ case is explained in the previous discussion itself in the name of $[L/D,d/D]=[1.4,0.2]$ variation. Hence, it is not discussed here as it is the common case. For the next $[d/D]$ variation at 0.26 and also for 0.34, the values of $[\Delta x_a/D]$ are observed to be at $\sim$0.1 until the critical $Re_D$ of $2.98 \times 10^5$. Above it, the values of $[\Delta x_a/D]$ jump only to $\sim$0.2, unlike the previous case ($[L/D,d/D]=[1.4,0.2]$). One of the reasons for a mild jump in $[\Delta x_a/D]$ is due to the thickness of the protrusion. It makes the separated shear layer to travel to a smaller length between the separation point and the forebody shoulder. Moreover, the flow deflection offered by the separated shear layer decreases considerably and alters the properties of the structures shed in the shear layer, especially at higher $Re_D$.

\begin{figure*}
  \centerline{\includegraphics[width=0.8\textwidth]{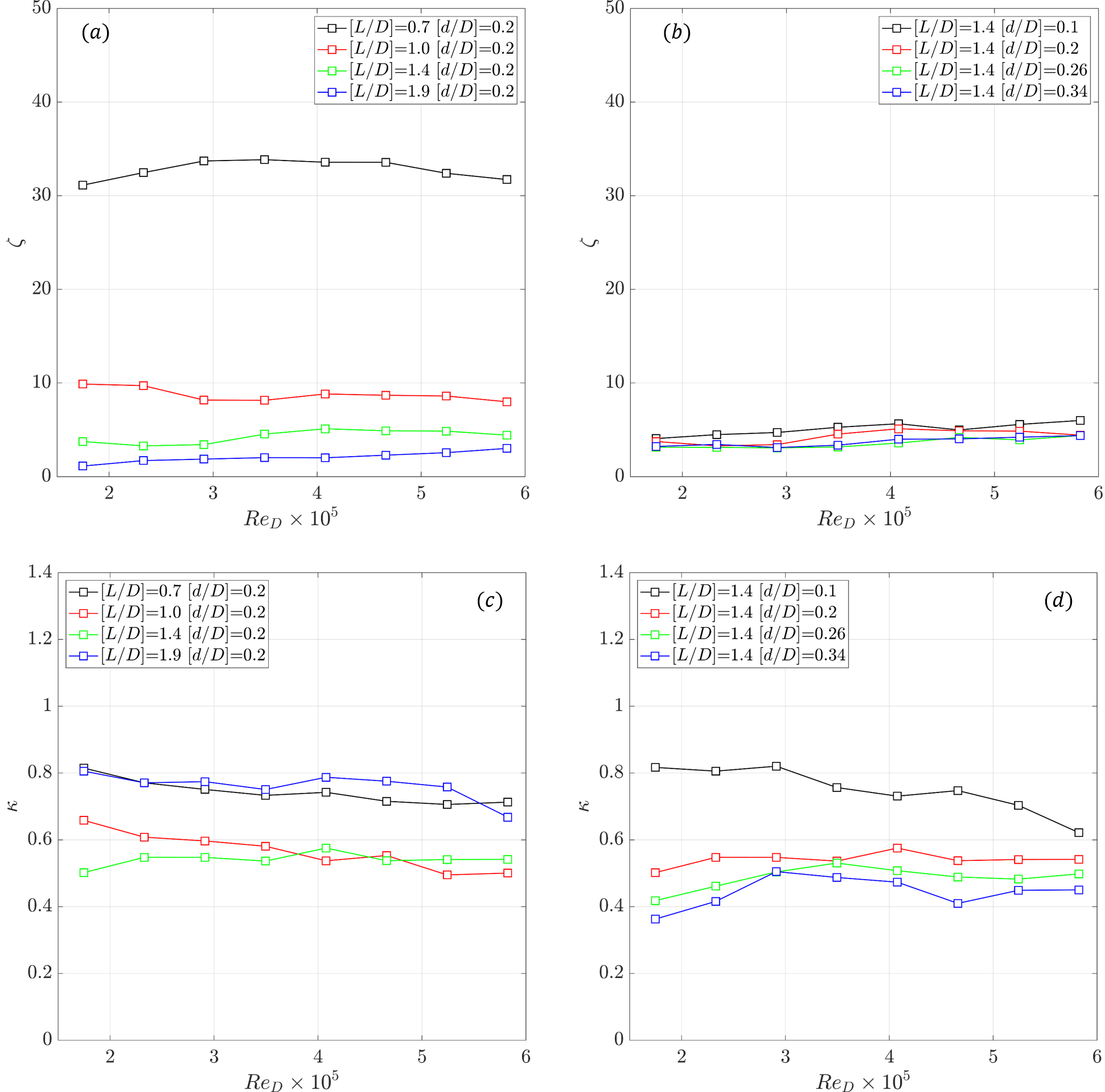}}
  \caption{Typical plots showing the variation of the (a-b) mean ($\zeta=\overline{p}/p_\infty$) and (c-d) fluctuating ($\kappa=\sqrt{\overline{p'^2}}/\overline{p}$) pressure while varying the (a, c) protrusion length-$L$ and the (b, d) protrusion diameter-$d$. The unsteady pressure data summarized here is taken at the $s_2$ location on the model's forebody as shown in Figure \ref{fig:model_description}.}
\label{fig:pressure_data}
\end{figure*}

\subsubsection{Apparent separated shear layer properties} \label{sec:shr_lyr_prop}
Rayleigh scattering images shown in Figure \ref{fig:plrs_mean_inst} are subjected to a certain image processing routines to extract the separated shock and shear layer properties. In particular, the operator based time-averaged image (bottom half of Figure \ref{fig:plrs_mean_inst}) are considered for their simplicity to draft the image processing algorithms. The processing routines are explained in Figure \ref{fig:shr_lyr_prop_fig}. One of the operator based time-averaged image is shown in Figure \ref{fig:shr_lyr_prop_fig}a. Grabbing intensity profile along the $y$-direction at two different $x$-locations are paramount for getting shock and shear layer properties. Quantities like shear layer deflection angle ($\theta$), shock angle ($\beta$), and shear layer growth rate ($\delta /x$) are computed closer to the forebody. A sample intensity profile at two arbitrary locations are depicted in Figure \ref{fig:shr_lyr_prop_fig}a as a solid red-color line. However, a more robust criteria (20\% and 30\% of the total $[L/D]$ in $[L/D]$ variations; 15\% and 35\% of the total $[L/D]$ in $[d/D]$ variations) is considered for extracting the actual intensity profiles from the Rayleigh scattering images. Intensity profile grabbing locations for each case are also marked in the bottom half of Figure \ref{fig:plrs_mean_inst} as dotted red-color line. The reason behind the selection of different criteria for $[L/D]$ and $[d/D]$ cases results from the necessity of grabbing a common location between the cases with turbulent shear layer in the vicinity. The shear layer properties are derived from a planar visualization image. Hence, the adopted processing routines yield only apparent values. However, the manner in which those values are acquired remain consistent across the whole $[L/D]$ and $[d/D]$ variations.

Between the obtained intensity profiles, the local peaks are observed to be growing linearly along the $x$-direction. After observing the operator based time-averaged image, such an assumption can be considered fairly a valid one. The manner through which the apparent values are extracted is depicted particularly in Figure \ref{fig:shr_lyr_prop_fig}b. The first peak is identified in the intensity profile by searching it from $[y/D] > 0$. Linearly connecting the peak $x-y$ locations from both the profiles result in calculating the line's slope which is basically the apparent shear layer deflection angle ($\theta$). Similarly, the second peak identification helps in marking the apparent separation shock angle ($\beta$). A half-width distance along the $y$-direction from the first peak identification in each profile is computed and represented as $\delta_1$ and $\delta_2$. As we already know the $x$-location in which these profiles are drawn, we can compute the apparent shear layer growth rate using the formula given below.
\begin{align}
    \left[ \frac{\delta}{x} \right] = \frac{|\delta_1 - \delta_2|}{|x_1 - x_2|}
\end{align}
All the computed apparent values are tabulated in Table \ref{tab:shr_lyr_prop}. In the tabulation, properties for $[L/D,d/D]=[0.7,0.2]$ are omitted as the shock is pulsating and no apparent shear layer properties can be extracted.

On the other hand, while increasing the $[L/D]$, a monotonically decreasing trend in all the shear layer properties is seen. As the separation point can be delayed for a longer $[L/D]$ and the separated shear layer needs to leave the forebody's shoulder tangentially, $\theta$ is observed to be less. Moreover, it also leads to the delayed transition from a laminar to a turbulent state in the separated shear layer. Such an event in-turn reduces the shear layer growth rate ($\delta/x$) drastically by $\sim$87\% between $[L/D]=1$ and 1.9. In contrast, there is no distinguishable trend that is seen in the apparent shear layer properties, while varying $[d/D]$. Most of the values are very close to each other and falls within the imaging uncertainty ($\pm 5\%$ in $\theta$ and $\beta$). Hence, for most of the discussion, the apparent shear layer properties can be considered almost a constant.  

\begin{figure*}
  \centerline{\includegraphics[width=\textwidth]{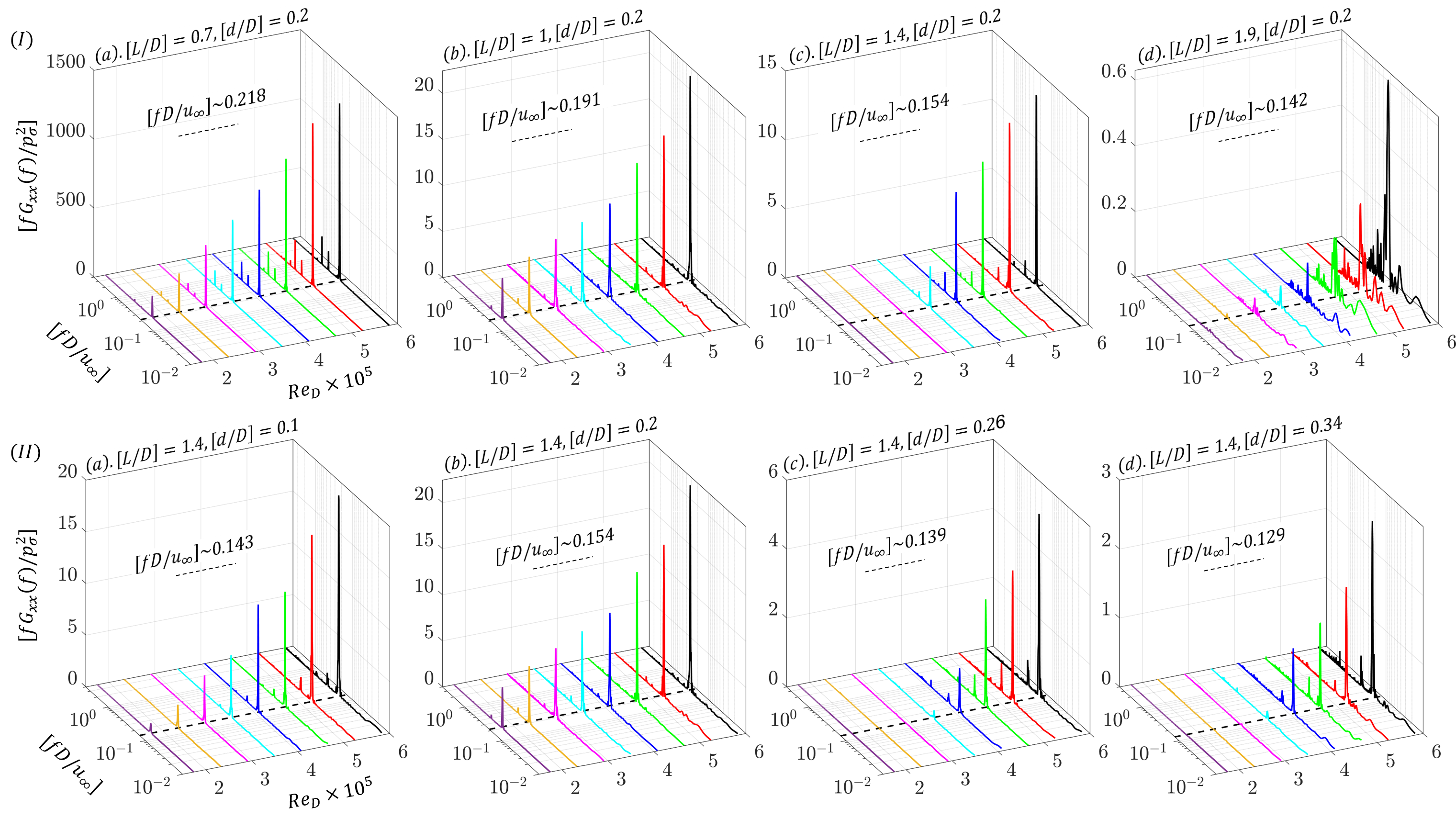}}
  \caption{Typical plots showing the variation of unsteady pressure's power spectra for different $Re_D$ emphasizing the effects of (I, a-d) protrusion length-$L$ and (II, a-d) protrusion diameter-$d$. The unsteady pressure spectra summarized here is taken at the $s_2$ location on the model forebody as shown in Figure \ref{fig:model_description}. The power spectra [$fG_{xx}(f)$] is normalized by the square of a reference pressure, $p_\sigma=1.45$ mPa. The dashed line in each plot represents the average [$fD/u_\infty$] observed across the different $Re_D$.}
\label{fig:unsteady_pressure}
\end{figure*}

\subsubsection{Pressure loading and fluctuation intensity}

The unsteady pressure transducer response located at a radial location of $2D/3$ (Figure \ref{fig:model_description}) is considered to access the pressure loading ($\zeta$) and fluctuation intensity ($\kappa$). The definition of these terms are given in the equation below.
\begin{align}
    \zeta =& \frac{\overline{p}}{p_\infty},\\
    \kappa =& \frac{\sqrt{\overline{p'^2}}}{\overline{p}}.
\end{align}
Variation in $\zeta$ and $\kappa$ on the forebody for different $[L/D]$ and $[d/D]$ cases at various $Re_D$ are plotted in Figure \ref{fig:pressure_data}. Values of $\zeta$ is obtained by taking the ratio of the average pressure ($\overline{p}$) during the useful run time and freestream static pressure ($p_\infty$). The variations of $\zeta$ for $[L/D]$ and $[d/D]$ cases are plotted in Figure \ref{fig:pressure_data} a-b. For the pulsating case ($[L/D,d/D]=[0.7,0.2]$), values of $\zeta$ lies almost a constant for the entire range of $Re_D$ at $\zeta \sim 32$. While increasing the $[L/D]$ to higher values, $\zeta$ is also observed to remain constant over the entire $Re_D$. At $[L/D]=1.0$, global $\zeta$ value remains around 9, which is $\sim$72\% smaller than the pulsating case. A further rise in $[L/D]$ to 1.4 and 1.9, results in the drop of $\zeta$ to 5 and 3, respectively. These values correspond to a total drop of $\sim$84\% and $\sim$91\% in global $\zeta$ with respect to the pulsating case. On contrary, while changing the $[d/D]$ across different $Re_D$, the global values of $\zeta$ remain almost a constant at 4. Thus, to reduce the mean pressure loading across a wide range of $Re_D$, the protrusion length needs to be increased.

Throughout the discussion in this paper, we highlighted the shock-related unsteadiness. Hence, understanding the fluctuation intensity ($\kappa$) due to shock and shear layer motions is of primary importance. A typical plot of $\kappa$ between different $[L/D]$, $[d/D]$ and $[Re_D]$ is given in Figure \ref{fig:pressure_data} c-d. For increasing $Re_D$, values of $\kappa$ are seen to be gradually reducing for most of the $[L/D]$ cases except $[L/D]=1.4$. The reduction is attributed to the stabilizing effect of the shear layer turbulence as explained in the previous section. For the pulsating flow ($[L/D,d/D]=[0.7,0.2]$), $\kappa$ is observed to be varying between 80\% to 70\%. While changing the $[L/D]$ to 1.0, these values are bounded between $0.65 \geq \kappa \geq 0.5$ which are considerably less than the previous case by $\sim$24\%. For $[L/D]=1.4$, transition in the shock oscillation happens at a critical $Re_D=2.98 \times 10^5$ as shown in the $x-t$ plots (Figure \ref{fig:xt_plot_all}-III a-h). As the sensors are located in the forebody, the pressure fluctuations detect two different types of motions across the critical $Re_D$. Hence, on a global sense, $\kappa$ value remains constant. A similar trend is seen in the cases of $[d/D]=[0.2,0.26,0.34]$ where such kind of transition occurs with a mean $\kappa$ varying about 50\%. While considering the longest protrusion case ($L/D=1.9$), values of $\kappa$ are observed to be higher as equivalent to the cases of pulsation across the entire $Re_D$ owing to the turbulent separated shear layer impinging on the forebody shoulder. For the thinnest protrusion ($d/D=0.1$), values of $\kappa$ bound between $0.82 \geq \kappa \geq 0.6$. As there are no transition in the pattern of shock motion across the $Re_D$, a drop in $\kappa$ is seen as before in the previous cases. In summary, while changing $[L/D]$, there exits a critical length up to which $\kappa$ value decreases. In the present case, the corresponding $[L/D]$ is 1.4 and the corresponding $\kappa$ is 0.5. On the other hand, while increasing the $[d/D]$, $\kappa$ values continues to drop. Hence, a thick protrusion is beneficial.

\subsubsection{Unsteady spectra}

Asides $\zeta$ and $\kappa$, computing the spectra from the unsteady pressure signal reveal the dominant unsteady components, if any. A series of spectra for different variations in $[L/D]$, $d/D$, and $Re_D$ is given in Figure \ref{fig:unsteady_pressure}. Effect of $[L/D]$ and $Re_D$ changes are particularly seen in Figure \ref{fig:unsteady_pressure}-I. For the shortest protrusion ($[L/D,d/D]=[0.7,0.2]$), a discrete non-dimensional frequency of $[fD/u_\infty] \sim 0.22$ is seen for all the $Re_D$ under consideration. The value is indeed matching with the $x-t$ plots shown in Figure \ref{fig:xt_plot_all}-I a-h. Noticeable variation is seen in the non-dimensional power content as the $Re_D$ changes. It varies almost linearly with $Re_D$ from 100 to 1200 (an increment of 12 times). The ramming of high freestream pressure on the forebody at high $Re_D$ is the reason behind the almost a linear variation in the non-dimensional power content. 

As the $[L/D]$ increases to 1, the non-dimensional frequency moves down to $[fD/u_\infty] \sim 0.19$. In the earlier discussion of $x-t$ plots (Figure \ref{fig:xt_plot_all}), we have already shown that the shock foot of the pulsation form of unsteadiness is packed with high-frequency sinuous wave movements, whereas the flapping one contains considerably a lower frequency with a saw-tooth wave. The non-dimensional power content of the discrete non-dimensional frequency varies between 5 to 20 (an increment of 4 times) across different $Re_D$. For $[L/D]=1.4$, until the critical $Re_D$ of $2.98\times 10^5$, no distinguishable non-dimensional power contents are visible in the spectra. The presence of low-amplitude broad oscillations seen from the $x-t$ plots (Figure \ref{fig:xt_plot_all}-III a-h) concur that the unsteady pressure is completely driven by the oscillation of shock and shear layer. Across the critical $Re_D$, a discrete non-dimensional frequency owing to flapping is seen with $[fD/u_\infty]\sim 0.15$. The non-dimensional power is seen to be varying between 5 to 15 (an increment by 3 times) across the $Re_D$. For the longest protrusion $[L/D]=1.9$, at lower $Re_D$ a weak non-dimensional spectra is seen with a non-dimensional power lesser than 0.1. However, as $Re_D$ increases, a discrete non-dimensional frequency begins to appear ($fD/u_\infty \sim 0.14$) and gradually turns broader. The impingement of turbulent structures from the separated shear layer on the forebody leads to the detection of unsteady pressure signal with broader spectral contents. In summary, a four order reduction in the non-dimensional power is seen while almost tripling the protrusion's length with the dominant non-dimensional spectra dropping by 0.6 times.

While changing the $[d/D]$ across $Re_D$ as shown in Figure \ref{fig:unsteady_pressure}-II, a similar pattern as seen in the $[L/D]$ cases is observed. The non-dimensional power content reduces by an order while tripling the $[d/D]$. The non-dimensional frequency however varies in a non-linear manner. Initially $[fD/u_\infty] \sim 0.14$ for $[d/D]=0.1$. While increasing the $[d/D]$ to 0.2, $[fD/u_\infty]$ raises to $\sim$0.15. From there on-wards, any further increment in $[d/D]$ (from 0.2 to 0.34) leads to an almost linear decay in the non-dimensional spectra from $[fD/u_\infty]=0.15$ to 0.13. The intermittent rise in $[fD/u_\infty]$ can be attributed to the formation of the separation point upstream to the expansion corner on the protrusion. Another noticeable feature in the unsteady pressure spectra is the presence of strong harmonics across all the cases of $[d/D]$. The non-dimensional power of these harmonics are in the comparable order of the fundamental. A strong resonating behavior in the shock/shear layer oscillation might be the reason to see the harmonics across the $[d/D]$ variations for this particular $[L/D]$ of 1.4.  

\subsubsection{Modal analysis}

\begin{figure*}
  \centerline{\includegraphics[width=0.8\textwidth]{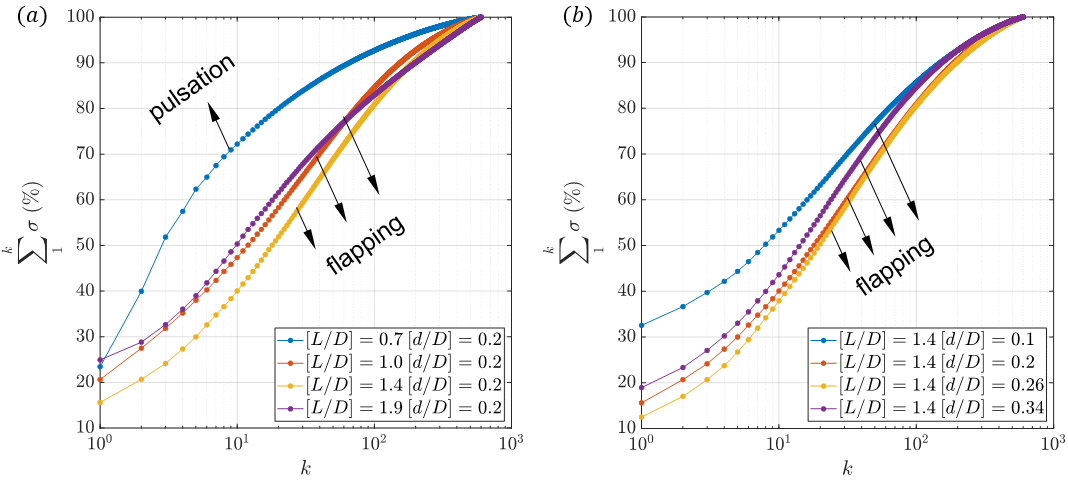}}
  \caption{Cumulative energy distribution from the Proper Orthogonal Decomposition (POD) of the Rayleigh scattering images using type-II imaging for variations in (a) $[L/D]$ and (b) $[d/D]$ across all the modes at $Re_D=5.85 \times 10^5$. The cases corresponding to pulsating and flapping type of shock/shear layer unsteadiness are marked for clarity.}
\label{fig:energy_plots}
\end{figure*}

Modal analysis \citep{Taira2017} like the Proper Orthogonal Decomposition (POD) reveal the dominant energetic contents that constitute the flow field both spatially and temporally. High-speed images obtained through several visualization techniques are subjected to POD analysis to extract the underlying dominant modes. In the present discussion, Rayleigh scattering images from type-II imaging as shown in Figure \ref{fig:plrs_mean_inst} are subjected to POD analysis. Rayleigh scattering images contain the distribution of a passive scalar across the flow domain. In a manner, the passive scalar indicates the variations in local density. Unsteadiness in the leading-edge separation problem at hypersonic speed is driven primarily by the shock and shear layer. Those components are well marked in a particular plane in terms of density variations using Rayleigh scattering technique. Analysing the Rayleigh scattering images using POD technique is called as snapshot POD. The images are generally stacked as a column matrix, and a covariance matrix is sought. Later, the matrix is decomposed as an eigenvalue problem, and the respective eigenvalues along with the eigenmodes are extracted. The eigenvalues are then ranked in the descending order of energy contents, and so the eigenmodes, to identify the dominant spatial mode. The time coefficient matrix is later constructed using a series of algebraic operations whose spectra reveal the dominant temporal contents in the flow. More details on the POD techniques are available in the review of \cite{Taira2017}.

The cumulative energy contents reveal the presence of the dominant modes across different $[L/D]$ and $[d/D]$ as shown in Figure \ref{fig:energy_plots}. As the shortest protrusion case ($[L/D,d/D]=[0.7,0.2]$) is driven by the dominant three phases of a typical pulsation cycle, the cumulative energy contents rise up quickly across a few modes. Half of the flow's total energy is just represented by the first three modes themselves. As the $[L/D]$ increases between $1 \leq [L/D] \leq 1.9$, the total number of modes required to represent half of the flow's total energy varies as $k=10$, $k=12$, and $k=18$, respectively. The requirement of larger number of modes to represent 50\% of the flow's total energy indicates the presence of multitude of significant modes. Multitude of modes arise in general due to flow turbulence as in jets or separated flows where the flow is dominated by the coupling of many events unlike in the case of discrete pulsation. Thus, for the flapping case, these mode numbers increases with $[L/D]$. However, the leading mode ($k=1$), especially for the longest protrusion ($[L/D]=1.9$) shows a drastic jump in energy to 25\%. The reason is attributed to the interaction of the turbulent separated shear layer with the forebody during flapping with a pronounced intensity fluctuation.

While varying the $[d/D]$, the thinnest protrusion contains higher energy contents in the leading modes. It progressively decays with increase in $[d/D]$ until 0.26. Above $[d/D]=0.26$, an energy increment is seen in the leading modes. The separation point in this case is placed very close to the protrusion tip as can be seen using Figure \ref{fig:plrs_mean_inst} II-d. The presence of downstream expansion corner and the varying recirculation zone might be attributed to the deviant behavior of the leading modes. It is also the reason behind the trend change in the total number of modes to represent half of the flow's total energy, as it varies as 8, 18, 20 and 14 while changing $[d/D]$ between $0.1 \leq [d/D] \leq 0.34$.

\begin{figure*}
  \centerline{\includegraphics[width=\textwidth]{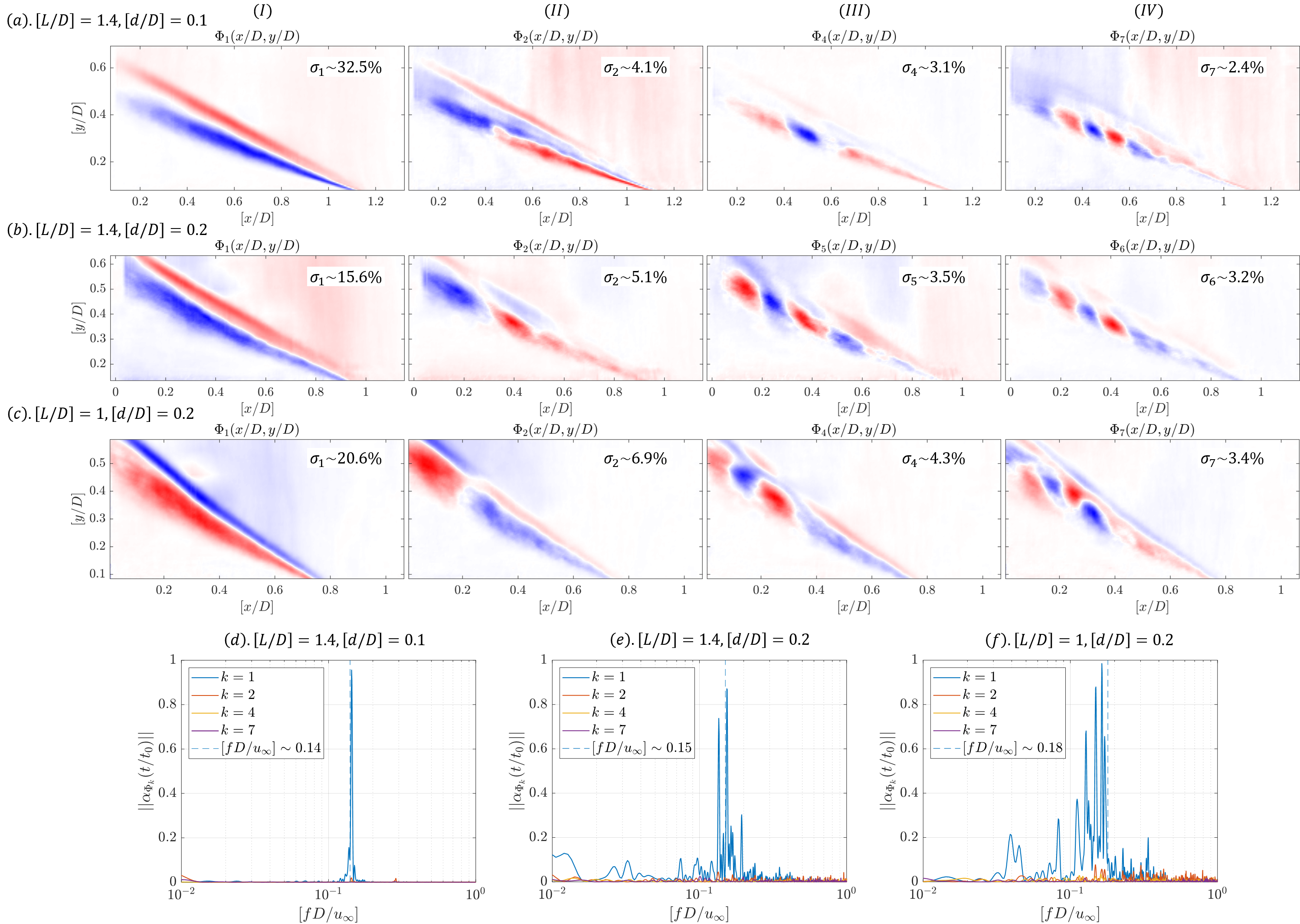}}
  \caption{(a-c) Contour plots showing the first four dominant spatial modes from the POD analysis of the Rayleigh scattering images using type-II imaging for three specific $[L/D]$ and $[d/D]$ cases at $Re_D-5.85 \times 10^5$; (e-f) Line plots showing the temporal spectra of the corresponding cases from the POD analysis.}
\label{fig:spatiotemporal_plots}
\end{figure*}

Identifying the leading spatial modes demand a careful investigation of the process and tuning the manner in which an image is acquired. In the recent work of \cite{Rao2019}, the influence of frame rate, and image exposure on the leading modes are explained using a screeching supersonic jet. While decomposing the images of flapping unsteadiness, the leading modes in all the $[L/D]$ and $[d/D]$ cases look the same. However, the order of appearance of the secondary spatial modes and the corresponding energy contents are different. In Figure \ref{fig:spatiotemporal_plots} a-c, the first four dominant spatial modes are shown for three different cases: one uncommon case from each $[L/D]$ and $[d/D]$, and the common case between them, for brevity. The first mode indicates the presence of separated shock and shear layer in alternating color. The color map (red-white-blue) indicates a correlation value between -1 to 1. Hence, the mode can be interpreted as events in the shock and shear layer region happening in out-of-phase. 
Physically, from the time-averaged or instantaneous Rayleigh scattering image (Figure \ref{fig:plrs_mean_inst}, see also the \href{https://youtu.be/JVMmbYXe0Pg}{supplementary} video), the variations in shock ($\beta$) and shear layer ($\theta$) angle can be seen. One can show that for a given upstream Mach number ($M$), the values of $\beta-\theta$ decreases when $\theta$ increases and vice versa using the $\theta-\beta-M$ relations for the conical flow \citep{sims1964tables}. During flapping, values of $\theta$ continuously change and using the $x-t$ plots and unsteady pressure monitors one can obtain the dominant spectra for such variations (Figure \ref{fig:xt_plot_all} and Figure \ref{fig:unsteady_pressure}). Thus, the first spatial mode can be interpreted as the coupled shock and shear layer mode that drives flapping about the separation point or simply, translatory flapping mode. The leading mode for the represented three cases contain 32.5\%, 15.6\%, and 20.6\% of the total flow energy. The corresponding temporal mode spectra shown in Figure \ref{fig:spatiotemporal_plots} d-f for $k=1$, reveal the dominant non-dimensional frequency lying close to the values as identified in Figure \ref{fig:xt_plot_all} and Figure \ref{fig:unsteady_pressure}. The only contrast is the broadband of spectra seen in the temporal plots of Figure \ref{fig:spatiotemporal_plots} d-f which can be attributed to the coupling of the leading mode with a few trailing modes in that specific case.

\begin{figure*}
  \centerline{\includegraphics[width=\textwidth]{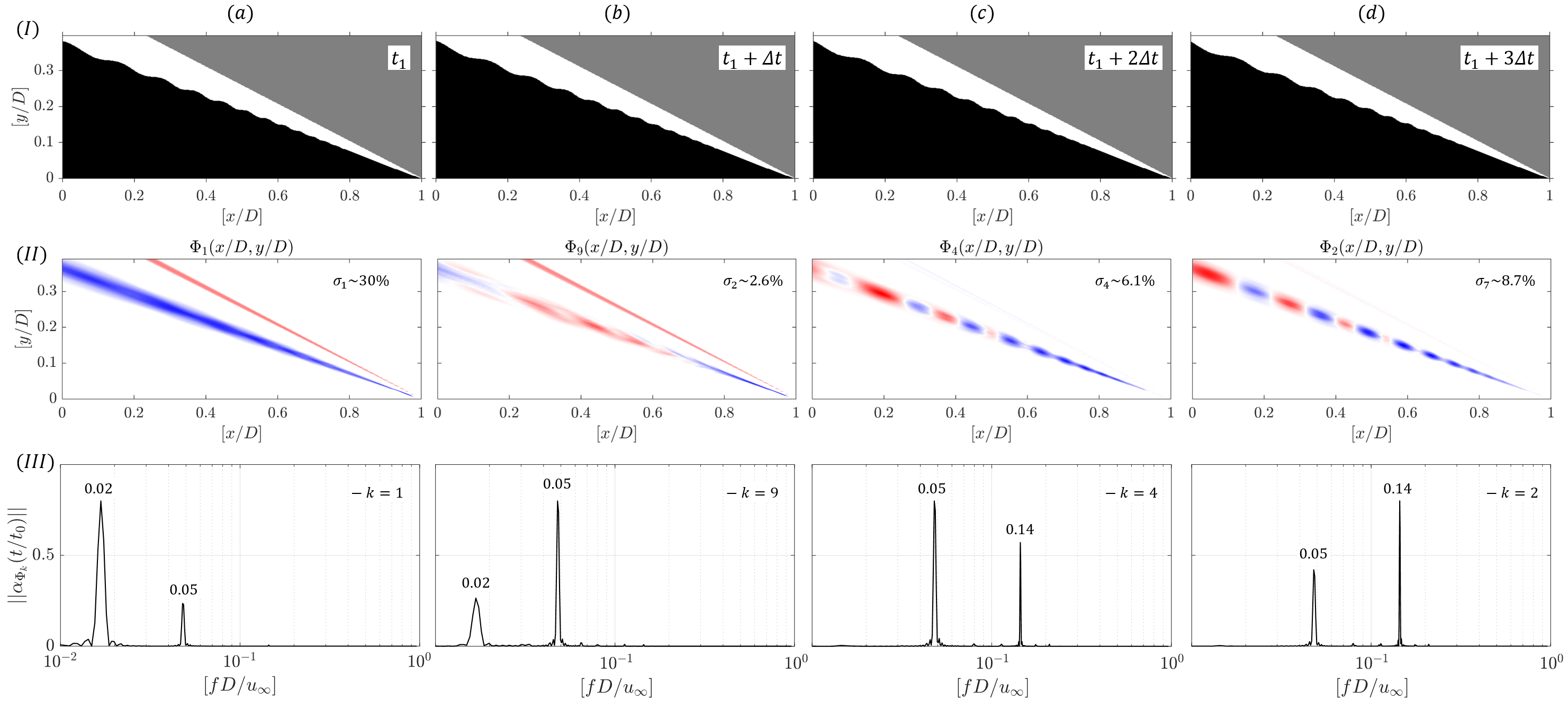}}
  \caption{(I a-d) Synthetic Rayleigh scattering image instants showing an equivalent leading-edge separation in a hypersonic flow. Corresponding video is also available in the \href{https://youtu.be/z6vRkbz2tnk}{supplementary}. (II a-d) Some of the vital energetic spatial modes from the POD analysis depicting translatory flapping, sinuous flapping, large-scale shedding, and small-scale shedding are given along with the corresponding energy contents. (III a-d) Line plot showing the variations in the non-dimensional temporal spectra across the different modes.}
\label{fig:synth_img_modes}
\end{figure*}

\begin{figure*}
  \centerline{\includegraphics[width=\textwidth]{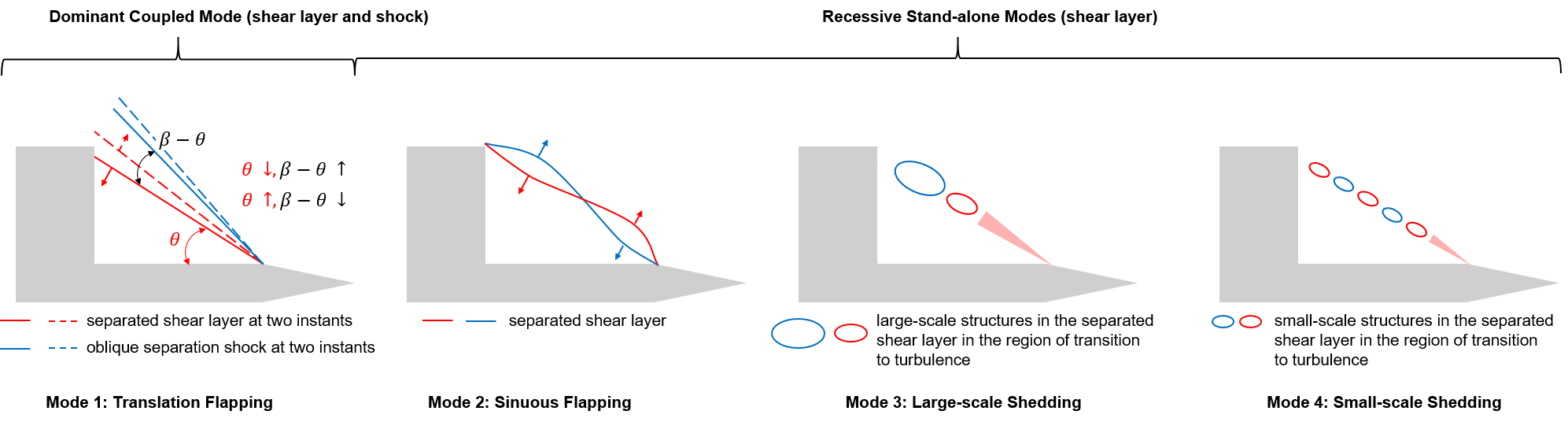}}
  \caption{A series of schematics representing the interpretation of the first four dominant spatial modes observed from the POD analysis of the Rayleigh scattering images and the synthetic images as shown in Figure \ref{fig:spatiotemporal_plots} and \ref{fig:synth_img_modes}.}
\label{fig:mode_schema}
\end{figure*}

The secondary modes are only composed of the separated shear layer, in particular, the sinuous flapping of the shear layer and the structures that are shed along its length. These convecting structures appear sometimes in mode pairs which are generally seen in the compressible jet or shear flows \cite{Rao2019,Rao_2020,Nanda_2021}. For the upcoming discussion, the best represented mode pairs are discussed in succession. In the selected $[L/D]$ and $[d/D]$ cases, the second mode is always sinuous flapping. The energy contents in the second mode is small in comparison with the leading mode (only in the range of 4\% to 7\%). The flapping spatial mode is identified by the presence of two distinct alternate color patch in the shear layer. The third and forth spatial modes contain the coherent structures present in the separated shear layer after it trips to a turbulent state. The third mode particularly contains large-scale structures while the fourth mode has considerably small-scale structures. The energy contents of these structures are very small compared to the leading mode. Across the considered cases, the third and the fourth mode have $\sim$3-4\% and $\sim$2-3\%, respectively. The respective temporal spectra shown in Figure \ref{fig:spatiotemporal_plots} d-f, do not contain any significant amplitude in comparison with the leading modes and they are mostly broadband.

However, making a call that a spatial mode as translatory flapping, sinuous flapping, or shedding is not direct. A separate modelling exercise is done using synthetic image sequences by representing most of the flow physics encountered in the leading-edge separation. Rayleigh scattering images are used as references to construct these synthetic images. Oblique separation shock is modelled as a line with higher slope than the separated shear layer. Shear layer is modelled as a line with amplifying sine wave whose wave number is decreasing along its length. A discrete frequency is given to simulate the flapping oscillation about the separation point and to create a travelling waveform along the shear layer. Equations responsible for the spatiotemporal modelling to simulate these synthetic images are given below. Within the spatial points of $0 \leq [x/D] \leq 3$ and temporal points of $1 \leq [t/t_0] \leq 25$ (where $t_0=1$ ms) at $f_s=40$ kHz, the oblique shock is modelled as,
\begin{align}
\label{eq:ob_eqn}
y_1 =& (1-0.8h)x,
\end{align}
where $h$ varies as,
\begin{align*}
    h = \lVert \sin (2\pi f_1 t) \rVert^{b_2}_{b_1}.
\end{align*}
The shear layer angle along with flapping and convection of transitional waves along its length are modeled as,
\begin{align}
\label{eq:shk_eqn}
    \begin{split}
        y_2 &= hx + a_1e^{-a_2/x} \cos  \left(\frac{a_3}{x^{a_4}+a_5} - 2\pi f_2t \right) ...\\ 
        &+ a_6e^{-a_7/x} \sin \left(\frac{a_8}{x^{a_9}+a_10} + 2\pi f_3t \right).
    \end{split}
\end{align}

In Eqn. \ref{eq:shk_eqn}, the first term represents the shock angle, the second term dictates the sinuous flapping about the shear layer, and the last term indicates the convection of instability or transitional waves along the shear layer. A synthetic image is prepared by shading the zones above Eqn. \ref{eq:ob_eqn} as moderate gray color to represent the low density passive scalar seeding in the freestream. Similarly, the zone bounded between Eqn. \ref{eq:ob_eqn} and \ref{eq:shk_eqn} is shaded in light gray color to represent the density jump across the separated shock. Region below Eqn. \ref{eq:shk_eqn} is colored as dark gray color as it is a recirculation zone. The overall color scheme is selected in a manner that it represents a typical Rayleigh scattering image. The constants value represented in Eqns. \ref{eq:ob_eqn}-\ref{eq:shk_eqn} are varied slightly to change the order of the resulting secondary modes. The constants considered for the present exercise is given as: $a_1$=0.5, $a_2$=10, $a_3$=0.1, $a_4$=-1, $a_5$=0, $a_6$=1, $a_7$=20, $a_8$=2000, $a_9$=1.2, $a_10$=20, $b_1$=0.35, $b_2$=0.38, $f_1$=350, $f_2$=1000, and $f_3$=3000. In Figure \ref{fig:synth_img_modes}, the generated synthetic image using the equations and the shading procedures mentioned above is shown. The time-resolved synthetic images are stitched together and represented in the \href{https://youtu.be/z6vRkbz2tnk}{supplementary} video to visualize the simulated shock and shear layer oscillation. The leading and secondary spatiotemporal modes obtained while using Eqns. \ref{eq:ob_eqn}-\ref{eq:shk_eqn} are shown in Figure \ref{fig:synth_img_modes}. As these modes are merely a representation of the underlying assumptions modelled in Eqn. \ref{eq:ob_eqn}-\ref{eq:shk_eqn}, the interpretation of the actual spatial modes in Figure \ref{fig:spatiotemporal_plots} a-c is considered plausible. A typical schematic shown in Figure \ref{fig:mode_schema} summarizes the realized spatial modes both from the actual Rayleigh scattering experiments and the synthetic images encountered in the leading-edge separation problems at hypersonic speed.

\section{Conclusions} \label{sec:conclusions}
Experiments to understand the leading-edge separation in hypersonic flows, particularly at $M_\infty=6$, are done using the newly commissioned Hypersonic Ludwieg Tunnel (HLT) at Technion for a wide range of Reynolds number between $1.66 \times 10^5 \leq Re_D \leq 5.85 \times 10^5$. Leading-edge separation is induced using an axial sharp-tip protrusion from a flat-face cylindrical forebody of diameter, $D=35$ mm. The length and diameter of the axial protrusion is changed as $[L/D]=[0.7,1.0,1.4,1.9]$ and $[d/D]=[0.1,0.2,0.26,0.34]$, respectively. Two visualization techniques like the schlieren imaging (density gradients along the transverse direction), and Rayleigh scattering are employed to understand the inherent flow structures responsible for the unsteadiness. Unsteady pressure transducers are used to extract the total pressure loading ($\zeta$), fluctuation intensity ($\kappa$), and a dominant power spectra for the different cases. Following are some of the major conclusions from the present study.
\begin{itemize}
    
\item From the schlieren imaging, the presence of two form of shock and shear layer unsteadiness are seen: pulsation (well-known and understood) and flapping (lesser-known and unclear). The $x-t$ and $x-f$ diagrams reveal pulsation to be the dominant one with high frequency sinuous shock motion all along the protrusion's length. The flapping case has considerably lesser amplitude, frequency and spatially confined within the protrusion's length with a saw-tooth type of shock motion. 
    
\item Except for the shortest protrusion length, all the other cases of $[L/D]$ and $[d/D]$ exhibit flapping form of unsteadiness. With increase in $[L/D]$ and $[d/D]$, flapping intensity decreases. At low $Re_D$ flapping cases exhibit a very low unsteadiness owing to the laminar separated shear layer. At high $Re_D$, the fluctuation intensity is high owing to the turbulent transition of the separated shear layer. The turbulent structures radiate a lot of noise in the recirculation region bounded by the forebody, protrusion, and shear layer.
    
\item From the Rayleigh scattering images, the presence of unstable shedding structures in the separated shear layer are seen at the highest $Re_D$ across all the cases of $[L/D]$ and $[d/D]$. The point of transition between a laminar to a turbulent state are seen to be varying along with the intensity of flapping. The properties of the unstable shear layer are extracted using a series of image processing routines. As $[L/D]$ varies, shock angle ($\beta$) and shear layer angle ($\theta$) monotonically decreases. In the $[d/D]$ cases, they almost remain the same.
    
\item From the unsteady pressure data, $\zeta$ remain almost the same across $Re_D$ and $\kappa$ gradually decreases, for all the cases of $[L/D]$ and $[d/D]$. Pulsation exhibits $\zeta$ closer to a value of 30, and decreases significantly with $[L/D]$ variations. Values of $\zeta$ remain almost invariant with changes in $[d/D]$. Values of $\kappa$ is observed to be around 80\% for the pulsating case and decreases gradually until for the longest protrusion case. Moreover, the separated shear layer undergoes transition to turbulence and hence, a sudden increment in $\kappa$ is seen for $[L/D]=1.9$. With increment in $[d/D]$, $\kappa$ gradually decreases.

\item The unsteady pressure spectra indicates a four order reduction in the non-dimensional power as the protrusion's length is almost tripled. Similarly, an order reduction in the non-dimensional power is seen while tripling the protrusion's diameter. Flapping produces a discrete spectra that decreases with $[L/D]$ variations. Except, some intermittency, the trend is seen to be similar across $[d/D]$ variations.

\item Modal analysis using Proper Orthogonal Decomposition (POD) of the Rayleigh scattering images reveals the presence of four driving modes in the flapping unsteadiness: translatory flapping, sinuous flapping, large-scale shedding, and small-scale shedding. The energy contents of these modes are seen to be varying based upon the intensity of flapping across $[L/D]$ and $[d/D]$ cases. However, in all of them, the leading mode remain the same with some rank changes between the secondary modes. The underlying modes are verified by simulating a series of synthetic images representing Rayleigh scattering and subjecting them to POD analysis.
\end{itemize}

\section*{Supplemental material}
The online version contains the following supplementary materials: 1. \href{https://youtu.be/AroYgvT9OCs}{video} of the high-speed schlieren imaging using type-II imaging, 2. \href{https://youtu.be/JVMmbYXe0Pg}{video} of the high-speed Rayleigh scattering imaging using type-II imaging, and 3. \href{https://youtu.be/z6vRkbz2tnk}{video} of the synthetic Rayleigh scattering imaging.

\section*{Acknowledgement}
The authors acknowledge the funding from the Israel Ministry of Defense to pursue the problem. The first and second authors are thankful to the partial financial help of Technion during their Post-Doctoral studies. The authors would like to thank the help of Mark Koifman, Michael Dunaevsky, Oleg Kan, Nadav Shefer, David Naftali, and Efim Shulman during the commissioning and testing of the Hypersonic Ludwieg Tunnel (HLT) at Technion Wind Tunnel Complex, effectively. All the authors have contributed equally to this paper. The authors report no conflict of interest.

\section*{Data availability statement}
The data that support the findings of this study are available from the corresponding author upon reasonable request.


\bibliographystyle{apalike}
\bibliography{references}

\begin{thebibliography}{}

\bibitem[Ahmed and Qin, 2011]{Ahmed2011}
Ahmed, M. and Qin, N. (2011).
\newblock Recent advances in the aerothermodynamics of spiked hypersonic
  vehicles.
\newblock 47(6):425--449.

\bibitem[Chung et~al., 2018]{chung_2018}
Chung, J.~D., Houim, R.~W., and Laurence, S.~J. (2018).
\newblock Theoretical and numerical study of a preheated ludwieg tube with
  adiabatic compression.
\newblock 56(10):3951--3962.

\bibitem[Coleman and Steele, 2009]{Coleman2009}
Coleman, H.~W. and Steele, W.~G. (2009).
\newblock {\em Experimentation, Validation, and Uncertainty Analysis for
  Engineers}.
\newblock John Wiley {\&} Sons, Inc.

\bibitem[Devaraj et~al., 2020]{Devaraj2020}
Devaraj, M. K.~K., Jutur, P., Rao, S. M.~V., Jagadeesh, G., and Anavardham, G.
  T.~K. (2020).
\newblock Experimental investigation of unstart dynamics driven by subsonic
  spillage in a hypersonic scramjet intake at mach 6.
\newblock {\em Physics of Fluids}, 32(2):026103.

\bibitem[Do et~al., 2010]{Do_2010}
Do, H., kyun Im, S., Mungal, M.~G., and Cappelli, M.~A. (2010).
\newblock Visualizing supersonic inlet duct unstart using planar laser rayleigh
  scattering.
\newblock {\em Experiments in Fluids}, 50(6):1651--1657.

\bibitem[Feszty et~al., 2004a]{Feszty2004b}
Feszty, D., Badcock, K.~J., and Richards, B.~E. (2004a).
\newblock Driving mechanism of high-speed unsteady spiked body flows, part 2:
  Oscillation mode.
\newblock {\em {AIAA} Journal}, 42(1):107--113.

\bibitem[Feszty et~al., 2004b]{Feszty2004a}
Feszty, D., Badcock, K.~J., and Richards, B.~E. (2004b).
\newblock Driving mechanisms of high-speed unsteady spiked body flows, part i:
  Pulsation mode.
\newblock {\em {AIAA} Journal}, 42(1):95--106.

\bibitem[Juliano et~al., 2008]{juliano_2008}
Juliano, T.~J., Schneider, S.~P., Aradag, S., and Knight, D. (2008).
\newblock Quiet-flow ludwieg tube for hypersonic transition research.
\newblock 46(7):1757--1763.

\bibitem[Karthick et~al., 2017]{Karthick_2017}
Karthick, S.~K., Rao, S. M.~V., Jagadeesh, G., and Reddy, K. P.~J. (2017).
\newblock Passive scalar mixing studies to identify the mixing length in a
  supersonic confined jet.
\newblock {\em Experiments in Fluids}, 58(5).

\bibitem[Kenworthy, 1978]{Kenworthy1978}
Kenworthy, M. (1978).
\newblock A study of unstable axisymmetric separation in high speed flows.
\newblock Ph.D. Dissertation, Dept. of Aerospace and Ocean Engineering,
  Virginia Polytechnic Inst. and State Univ.

\bibitem[Kulkarni and Reddy, 2008]{Kulkarni2008}
Kulkarni, V. and Reddy, K. P.~J. (2008).
\newblock Enhancement in counterflow drag reduction by supersonic jet in high
  enthalpy flows.
\newblock {\em Physics of Fluids}, 20(1):016103.

\bibitem[Labuda et~al., 2020]{labuda_2020}
Labuda, D., Komives, J., Reeder, M.~F., Borg, M.~P., and Jewell, J.~S. (2020).
\newblock High-speed schlieren visualization of mach 6 flow past a cone with
  varied parameters.
\newblock American Institute of Aeronautics and Astronautics.

\bibitem[Li et~al., 2022]{Li_2022}
Li, Z., Xiong, Y., Yuan, X., Chen, J., Zhao, J., and Wu, J. (2022).
\newblock A-variant design of hypersonic ludwieg tube wind tunnel.
\newblock {\em {AIAA} Journal}, 60(7):3990--4005.

\bibitem[Maull, 1960]{Maull1960584}
Maull, D.~J. (1960).
\newblock Hypersonic flow over axially symmetric spiked bodies.
\newblock 8(04):584.

\bibitem[Nanda et~al., 2021]{Nanda_2021}
Nanda, S.~R., Karthick, S.~K., Krishna, T.~V., De, A., and Ibrahim, M.~S.
  (2021).
\newblock On the unsteady dynamics of partially shrouded compressible jets.
\newblock {\em Experiments in Fluids}, 62(10).

\bibitem[Neet and Austin, 2020]{neet_2020}
Neet, M.~C. and Austin, J.~M. (2020).
\newblock Effects of surface compliance on shock boundary layer interaction in
  the caltech mach 4 ludwieg tube.
\newblock American Institute of Aeronautics and Astronautics.

\bibitem[Panaras and Drikakis, 2009]{Panaras200969}
Panaras, A.~G. and Drikakis, D. (2009).
\newblock High-speed unsteady flows around spiked-blunt bodies.
\newblock 632:69--96.

\bibitem[Rao and Karthick, 2019]{Rao2019}
Rao, S.~M. and Karthick, S. (2019).
\newblock Studies on the effect of imaging parameters on dynamic mode
  decomposition of time-resolved schlieren flow images.
\newblock {\em Aerospace Science and Technology}, 88:136--146.

\bibitem[Rao et~al., 2020]{Rao_2020}
Rao, S. M.~V., Karthick, S.~K., and Anand, A. (2020).
\newblock Elliptic supersonic jet morphology manipulation using sharp-tipped
  lobes.
\newblock {\em Physics of Fluids}, 32(8):086107.

\bibitem[Russell and Tong, 1973]{Russell_1973}
Russell, D.~A. and Tong, K.-O. (1973).
\newblock Aerodynamics of high-performance ludwieg tubes.
\newblock {\em {AIAA} Journal}, 11(5):642--648.

\bibitem[Sahoo et~al., 2020]{sahoo_2020}
Sahoo, D., Karthick, S.~K., Das, S., and Cohen, J. (2020).
\newblock Parametric experimental studies on supersonic flow unsteadiness over
  a hemispherical spiked body.
\newblock 58(8):3446--3463.

\bibitem[Sahoo et~al., 2021]{sahoo_2021}
Sahoo, D., Karthick, S.~K., Das, S., and Cohen, J. (2021).
\newblock Shock-related unsteadiness of axisymmetric spiked bodies in
  supersonic flow.
\newblock 62(4).

\bibitem[Schneider, 1992]{Schneider_1992}
Schneider, S. (1992).
\newblock A quiet-flow ludwieg tube for experimental study of high speed
  boundary layer transition.
\newblock In {\em 28th Joint Propulsion Conference and Exhibit}. American
  Institute of Aeronautics and Astronautics.

\bibitem[Schrijer and Bannink, 2010]{Schrijer_2010}
Schrijer, F. F.~J. and Bannink, W.~J. (2010).
\newblock Description and flow assessment of the delft hypersonic ludwieg tube.
\newblock {\em Journal of Spacecraft and Rockets}, 47(1):125--133.

\bibitem[Sch\"{u}lein, 2004]{schulein_2004}
Sch\"{u}lein, E. (2004).
\newblock Optical skin friction measurements in short-duration facilities
  (invited).
\newblock American Institute of Aeronautics and Astronautics.

\bibitem[Segal, 2011]{segal_2011}
Segal, C. (2011).
\newblock A unique, mach 6 enthalpy, non-vitiated facility for hypersonic
  aerodynamics research.
\newblock American Institute of Aeronautics and Astronautics.

\bibitem[Sekar et~al., 2020]{Sekar2020}
Sekar, K.~R., Karthick, S.~K., Jegadheeswaran, S., and Kannan, R. (2020).
\newblock On the unsteady throttling dynamics and scaling analysis in a typical
  hypersonic inlet{\textendash}isolator flow.
\newblock {\em Physics of Fluids}, 32(12):126104.

\bibitem[Settles, 2001]{Settles2001}
Settles, G.~S. (2001).
\newblock {\em Schlieren and Shadowgraph Techniques}.
\newblock Springer Berlin Heidelberg.

\bibitem[Sims, 1964]{sims1964tables}
Sims, J.~L. (1964).
\newblock {\em Tables for supersonic flow around right circular cones at zero
  angle of attack}, volume 3004.
\newblock Office of Scientific and Technical Information, National Aeronautics
  and~….

\bibitem[Sugarno et~al., 2022]{Sugarno2022}
Sugarno, M.~I., Sriram, R., Karthick, S.~K., and Jagadeesh, G. (2022).
\newblock Unsteady pulsating flowfield over spiked axisymmetric forebody at
  hypersonic flows.
\newblock {\em Physics of Fluids}, 34(1):016104.

\bibitem[Taira et~al., 2017]{Taira2017}
Taira, K., Brunton, S.~L., Dawson, S. T.~M., Rowley, C.~W., Colonius, T.,
  McKeon, B.~J., Schmidt, O.~T., Gordeyev, S., Theofilis, V., and Ukeiley,
  L.~S. (2017).
\newblock Modal analysis of fluid flows: An overview.
\newblock {\em {AIAA} Journal}, 55(12):4013--4041.

\bibitem[Venukumar et~al., 2006]{Venukumar2006}
Venukumar, B., Jagadeesh, G., and Reddy, K. P.~J. (2006).
\newblock Counterflow drag reduction by supersonic jet for a blunt body in
  hypersonic flow.
\newblock {\em Physics of Fluids}, 18(11):118104.

\bibitem[Zhang and Lee, 2016]{Zhang_2016}
Zhang, C. and Lee, C. (2016).
\newblock Rayleigh-scattering visualization of the development of second-mode
  waves.
\newblock {\em Journal of Visualization}, 20(1):7--12.

\end{thebibliography}

\onecolumngrid

\PRLsep	

\end{document}